\documentclass[twocolumn]{aastex62}
\usepackage{xcolor,soul}
\providecommand{\bjdtdb}{\ensuremath{\rm {BJD_{TDB}}}}

\providecommand{\teff}{\ensuremath{T_{\rm eff}}}

\providecommand{\msun}{\ensuremath{\,M_\Sun}}
\providecommand{\rsun}{\ensuremath{\,R_\Sun}}
\providecommand{\lsun}{\ensuremath{\,L_\Sun}}
\providecommand{\mj}{\ensuremath{\,{\rm M_J}}}
\providecommand{\rj}{\ensuremath{\,{\rm R_J}}}

\providecommand{\mst}{\ensuremath{\,{\rm M_\odot}}}
\providecommand{\rst}{\ensuremath{\,{\rm R_\odot}}}

\providecommand{\arcsec}{$^{\prime \prime}$}

\graphicspath{{./}{figures/}}

\begin{document}

\title{TOI-503: The first known brown dwarf-Am star binary from the TESS mission
\footnote{This work is done under the framework of the KESPRINT
collaboration ({\tt{http://kesprint.science}}). KESPRINT is an international consortium devoted to the characterisation and research of exoplanets discovered with space-based missions.}}

\author[0000-0002-5313-9722]{J\'an \v{S}ubjak}
\affil{\rm Astronomical Institute, Czech Academy of Sciences, Fri{\v c}ova 298, 251 65, Ond\v{r}ejov, Czech Republic}
\affil{\rm Astronomical Institute of Charles University, V Hole\v{s}ovi{\v c}k\'ach 2, 180 00, Praha, Czech Republic}

\author[0000-0001-8983-5300]{Rishikesh Sharma}
\affil{\rm Astronomy \& Astrophysics Division, Physical Research Laboratory, Ahmedabad 380009, India}

\author[0000-0001-6416-1274]{Theron W. Carmichael}
\affil{\rm Harvard University,
Cambridge, MA 02138}
\affil{\rm Center for Astrophysics ${\rm \mid}$ Harvard {\rm \&} Smithsonian, 60 Garden Street, Cambridge, MA 02138, USA}
\affil{\rm National Science Foundation Graduate Research Fellow}

\author[0000-0002-5099-8185]{Marshall C. Johnson}
\affil{\rm Las Cumbres Observatory, 6740 Cortona Drive, Suite 102, Goleta, CA 93117, USA}

\author{Erica J. Gonzales}
\affil{\rm University of California, Santa Cruz, 1156 High Street, Santa Cruz CA 95065, USA}
\affil{\rm National Science Foundation Graduate Research Fellow}

\author{Elisabeth Matthews}
\affil{\rm Department of Physics, and Kavli Institute for Astrophysics and Space Research, Massachusetts Institute of Technology, Cambridge, MA 02139, USA}

\author[0000-0002-9486-4840]{Henri M. J. Boffin}
\affil{\rm ESO, Karl-Schwarzschild-Stra{\ss}e 2, 85748 Garching bei M\"unchen, Germany}

\author{Rafael Brahm}
\affil{\rm Center of Astro-Engineering UC, Pontificia Universidad Cat\'olica de Chile, Av. Vicuña Mackenna 4860, 7820436 Macul, Santiago, Chile}
\affil{\rm Instituto de Astrof\'{i}sica, Pontificia Universidad Cat\'olica de Chile, Av. Vicuña Mackenna 4860, Macul, Santiago, Chile}
\affil{\rm Millennium Institute for Astrophysics, Chile}

\author[0000-0002-1887-1192]{Priyanka Chaturvedi}
\affil{\rm Th\"uringer Landessternwarte Tautenburg, Sternwarte 5, 07778 Tautenburg, Germany}

\author[0000-0002-3815-8407]{Abhijit Chakraborty}
\affil{\rm Astronomy \& Astrophysics Division, Physical Research Laboratory, Ahmedabad 380009, India}

\author{David R. Ciardi}
\affil{\rm Caltech/IPAC-NASA Exoplanet Science Institute, M/S 100-22, 770 S. Wilson Ave, Pasadena, CA 91106, USA}

\author[0000-0001-6588-9574]{Karen A. Collins}
\affil{\rm Center for Astrophysics ${\rm \mid}$ Harvard {\rm \&} Smithsonian, 60 Garden Street, Cambridge, MA 02138, USA}

\author{Massimiliano Esposito}
\affil{\rm Th\"uringer Landessternwarte Tautenburg, Sternwarte 5, 07778 Tautenburg, Germany}

\author{Malcolm Fridlund}
\affil{\rm Chalmers University of Technology, Department of Space, Earth and Environment, Onsala Space Observatory, SE-439 92 Onsala, Sweden}
\affil{\rm Leiden Observatory, University of Leiden, PO Box 9513, 2300 RA, Leiden, The Netherlands}

\author{Tianjun Gan}
\affil{\rm Physics Department and Tsinghua Centre for Astrophysics, Tsinghua University, Beijing 100084, China}

\author[0000-0001-8627-9628]{Davide Gandolfi}
\affil{\rm Dipartimento di Fisica, Universit\`a degli Studi di Torino, via Pietro Giuria 1, I-10125, Torino, Italy}

\author{Rafael A. Garc\'ia}
\affil{\rm IRFU, CEA, Universit\'e Paris-Saclay, Gif-sur-Yvette, France}
\affil{\rm AIM, CEA, CNRS, Universit\'e Paris-Saclay, Universit\'e Paris Diderot, Sorbonne Paris Cit\'e, F-91191 Gif-sur-Yvette, France}

\author{Eike Guenther}
\affil{\rm \rm Th\"uringer Landessternwarte Tautenburg, Sternwarte 5, 07778 Tautenburg, Germany}

\author{Artie Hatzes}
\affil{\rm \rm Th\"uringer Landessternwarte Tautenburg, Sternwarte 5, 07778 Tautenburg, Germany}

\author{David W. Latham}
\affil{\rm Center for Astrophysics ${\rm \mid}$ Harvard {\rm \&} Smithsonian, 60 Garden Street, Cambridge, MA 02138, USA}

\author{St\'ephane Mathis}
\affil{\rm IRFU, CEA, Universit\'e Paris-Saclay, Gif-sur-Yvette, France}
\affil{\rm AIM, CEA, CNRS, Universit\'e Paris-Saclay, Universit\'e Paris Diderot, Sorbonne Paris Cit\'e, F-91191 Gif-sur-Yvette, France}

\author{Savita Mathur}
\affil{\rm Instituto de Astrof\'isica de Canarias, C/ V\'ia L\'actea s/n, E-38205 La Laguna, Spain}
\affil{\rm Departamento de Astrof\'isica, Universidad de La Laguna, E-38206 La Laguna, Spain}

\author{Carina M. Persson}
\affil{\rm Chalmers University of Technology, Department of Space, Earth and Environment, Onsala Space Observatory, SE-439 92 Onsala, Sweden}

\author{Howard M. Relles}
\affil{\rm Center for Astrophysics ${\rm \mid}$ Harvard {\rm \&} Smithsonian, 60 Garden Street, Cambridge, MA 02138, USA}

\author{Joshua E. Schlieder}
\affil{\rm Exoplanets and Stellar Astrophysics Laboratory, Code 667, NASA Goddard Space Flight Center, Greenbelt, MD 20771, USA}

\author{Thomas Barclay}
\affil{\rm NASA Goddard Space Flight Center, Greenbelt, MD 20771}

\author{Courtney D. Dressing}
\affil{\rm Astronomy Department, University of California, Berkeley, CA 94720, USA}

\author{Ian Crossfield}
\affil{\rm Department of Physics, and Kavli Institute for Astrophysics and Space Research, Massachusetts Institute of Technology, Cambridge, MA 02139, USA}

\author{Andrew W. Howard}
\affil{\rm California Institute of Technology, Pasadena, CA 91125, USA}

\author{Florian Rodler}
\affil{\rm European Southern Observatory, Alonso de C\'ordova 3107, Vitacura, Santiago, Chile}

\author{George Zhou}
\affil{\rm Center for Astrophysics ${\rm \mid}$ Harvard {\rm \&} Smithsonian, 60 Garden Street, Cambridge, MA 02138, USA}

\author[0000-0002-8964-8377]{Samuel N. Quinn}
\affil{\rm Center for Astrophysics ${\rm \mid}$ Harvard {\rm \&} Smithsonian, 60 Garden Street, Cambridge, MA 02138, USA}

\author{Gilbert A. Esquerdo}
\affil{\rm Center for Astrophysics ${\rm \mid}$ Harvard {\rm \&} Smithsonian, 60 Garden Street, Cambridge, MA 02138, USA}

\author{Michael L. Calkins}
\affil{\rm Center for Astrophysics ${\rm \mid}$ Harvard {\rm \&} Smithsonian, 60 Garden Street, Cambridge, MA 02138, USA}

\author{Perry Berlind}
\affil{\rm Center for Astrophysics ${\rm \mid}$ Harvard {\rm \&} Smithsonian, 60 Garden Street, Cambridge, MA 02138, USA}

\author[0000-0002-3481-9052]{Keivan G.\ Stassun}
\affil{Vanderbilt University, Department of Physics \& Astronomy, 6301 Stevenson Center Ln., Nashville, TN 37235, USA}
\affil{Fisk University, Department of Physics, 1000 18th Ave. N., Nashville, TN 37208, USA}

\author{Martin Bla{\v z}ek}
\affil{\rm Astronomical Institute, Czech Academy of Sciences, Fri{\v c}ova 298, 251 65, Ond\v{r}ejov, Czech Republic}
\affil{\rm Department of Theoretical Physics and Astrophysics, Masaryk University, Kotl\'a{\v r}sk\'a 2, 61137 Brno, Czech Republic}

\author[0000-0002-7602-0046]{Marek Skarka}
\affil{\rm Astronomical Institute, Czech Academy of Sciences, Fri{\v c}ova 298, 251 65, Ond\v{r}ejov, Czech Republic}
\affil{\rm Department of Theoretical Physics and Astrophysics, Masaryk University, Kotl\'a{\v r}sk\'a 2, 61137 Brno, Czech Republic}

\author{Magdalena {\v S}pokov\'a}
\affil{\rm Astronomical Institute, Czech Academy of Sciences, Fri{\v c}ova 298, 251 65, Ond\v{r}ejov, Czech Republic}
\affil{\rm Department of Theoretical Physics and Astrophysics, Masaryk University, Kotl\'a{\v r}sk\'a 2, 61137 Brno, Czech Republic}

\author{Ji{\v r}\'i {\v Z}\'ak}
\affil{\rm Department of Theoretical Physics and Astrophysics, Masaryk University, Kotl\'a{\v r}sk\'a 2, 61137 Brno, Czech Republic}

\author{Simon Albrecht}
\affil{\rm Stellar Astrophysics Centre, Department of Physics and Astronomy, Aarhus University, Ny Munkegade 120, DK-8000 Aarhus C, Denmark}

\author{Roi Alonso Sobrino}
\affil{\rm Instituto de Astrof\'isica de Canarias, C/ V\'ia L\'actea s/n, E-38205 La Laguna, Spain}
\affil{\rm Departamento de Astrof\'isica, Universidad de La Laguna, E-38206 La Laguna, Spain}

\author{Paul Beck}
\affil{\rm Instituto de Astrof\'isica de Canarias, C/ V\'ia L\'actea s/n, E-38205 La Laguna, Spain}
\affil{\rm Departamento de Astrof\'isica, Universidad de La Laguna, E-38206 La Laguna, Spain}
\affil{\rm Institut for Physics, Karl-Franzens University of Graz, Universit\"atsplatz 5, 8020 Graz, Austria}

\author{Juan Cabrera}
\affil{\rm Institute of Planetary Research, German Aerospace Center, Rutherfordstrasse 2, D-12489 Berlin, Germany}

\author{Ilaria Carleo}
\affil{\rm Astronomy Department and Van Vleck Observatory, Wesleyan University, Middletown, CT 06459, USA}

\author{William D. Cochran}
\affil{\rm Department of Astronomy and McDonald Observatory, University of Texas at Austin, 2515 Speedway, Stop C1400, Austin, TX 78712, USA}

\author{Szilard Csizmadia}
\affil{\rm Institute of Planetary Research, German Aerospace Center, Rutherfordstrasse 2, D-12489 Berlin, Germany}

\author{Fei Dai}
\affil{\rm Department of Physics and Kavli Institute for Astrophysics and Space Research, Massachusetts Institute of Technology, Cambridge, MA 02139, USA}
\affil{\rm Department of Astrophysical Sciences, Princeton University, 4 Ivy Lane, Princeton, NJ 08544, USA}

\author{Hans J. Deeg}
\affil{\rm Instituto de Astrof\'isica de Canarias, C/ V\'ia L\'actea s/n, E-38205 La Laguna, Spain}
\affil{\rm Departamento de Astrof\'isica, Universidad de La Laguna, E-38206 La Laguna, Spain}

\author{Jerome P. de Leon}
\affil{\rm Department of Astronomy, The University of Tokyo, 7-3-1 Hongo, Bunkyo-ku, Tokyo 113-0033, Japan}

\author{Philipp Eigm\"uller}
\affil{\rm Institute of Planetary Research, German Aerospace Center, Rutherfordstrasse 2, D-12489 Berlin, Germany}

\author{Michael Endl}
\affil{\rm Department of Astronomy and McDonald Observatory, University of Texas at Austin, 2515 Speedway, Stop C1400, Austin, TX 78712, USA}

\author{Anders Erikson}
\affil{\rm Institute of Planetary Research, German Aerospace Center, Rutherfordstrasse 2, D-12489 Berlin, Germany}

\author{Akai Fukui}
\affil{\rm National Astronomical Observatory of Japan, 2-21-1 Osawa, Mitaka, Tokyo 181-8588, Japan}

\author{Iskra Georgieva}
\affil{\rm Chalmers University of Technology, Department of Space, Earth and Environment, Onsala Space Observatory, SE-439 92 Onsala, Sweden}

\author{Luc\'ia Gonz\'alez-Cuesta}
\affil{\rm Instituto de Astrof\'isica de Canarias, C/ V\'ia L\'actea s/n, E-38205 La Laguna, Spain}
\affil{\rm Departamento de Astrof\'isica, Universidad de La Laguna, E-38206 La Laguna, Spain}

\author{Sascha Grziwa}
\affil{\rm Rheinisches Institut f\"ur Umweltforschung an der Universit\"at zu K\"oln, Aachener Strasse 209, D-50931 K\"oln, Germany}

\author{Diego Hidalgo}
\affil{\rm Instituto de Astrof\'isica de Canarias, C/ V\'ia L\'actea s/n, E-38205 La Laguna, Spain}
\affil{\rm Departamento de Astrof\'isica, Universidad de La Laguna, E-38206 La Laguna, Spain}

\author{Teruyuki Hirano}
\affil{\rm Department of Earth and Planetary Sciences, Tokyo Institute of Technology, 2-12-1 Ookayama, Meguro-ku, Tokio 152-8551, Japan}

\author{Maria Hjorth}
\affil{\rm Stellar Astrophysics Centre, Department of Physics and Astronomy, Aarhus University, Ny Munkegade 120, DK-8000 Aarhus C, Denmark}

\author{Emil Knudstrup}
\affil{\rm Stellar Astrophysics Centre, Department of Physics and Astronomy, Aarhus University, Ny Munkegade 120, DK-8000 Aarhus C, Denmark}

\author{Judith Korth}
\affil{\rm Rheinisches Institut f\"ur Umweltforschung an der Universit\"at zu K\"oln, Aachener Strasse 209, D-50931 K\"oln, Germany}

\author{Kristine W. F. Lam}
\affil{\rm Zentrum f\"ur Astronomie und Astrophysik, Technische Universit\"at Berlin, Hardenbergstr. 36, 10623 Berlin, Germany}

\author[0000-0002-4881-3620]{John H. Livingston}
\affil{\rm Department of Astronomy, The University of Tokyo, 7-3-1 Hongo, Bunkyo-ku, Tokyo 113-0033, Japan}

\author{Mikkel N. Lund}
\affil{\rm Stellar Astrophysics Centre, Department of Physics and Astronomy, Aarhus University, Ny Munkegade 120, DK-8000 Aarhus C, Denmark}

\author{Rafael Luque}
\affil{\rm Instituto de Astrof\'isica de Canarias, C/ V\'ia L\'actea s/n, E-38205 La Laguna, Spain}
\affil{\rm Departamento de Astrof\'isica, Universidad de La Laguna, E-38206 La Laguna, Spain}

\author{Pilar Montanes Rodr\'iguez}
\affil{\rm Instituto de Astrof\'isica de Canarias, C/ V\'ia L\'actea s/n, E-38205 La Laguna, Spain}
\affil{\rm Departamento de Astrof\'isica, Universidad de La Laguna, E-38206 La Laguna, Spain}

\author{Felipe Murgas}
\affil{\rm Instituto de Astrof\'isica de Canarias, C/ V\'ia L\'actea s/n, E-38205 La Laguna, Spain}
\affil{\rm Departamento de Astrof\'isica, Universidad de La Laguna, E-38206 La Laguna, Spain}

\author{Norio Narita}
\affil{\rm Astrobiology Center, 2-21-1 Osawa, Mitaka, Tokyo 181-8588, Japan}
\affil{\rm JST, PRESTO, 2-21-1 Osawa, Mitaka, Tokyo 181-8588, Japan}
\affil{\rm National Astronomical Observatory of Japan, 2-21-1 Osawa, Mitaka, Tokyo 181-8588, Japan}
\affil{\rm Instituto de Astrof\'isica de Canarias, C/ V\'ia L\'actea s/n, E-38205 La Laguna, Spain}

\author{David Nespral}
\affil{\rm Instituto de Astrof\'isica de Canarias, C/ V\'ia L\'actea s/n, E-38205 La Laguna, Spain}
\affil{\rm Departamento de Astrof\'isica, Universidad de La Laguna, E-38206 La Laguna, Spain}

\author{Prajwal Niraula}
\affil{\rm Department of Earth, Atmospheric and Planetary Sciences, MIT, 77 Massachusetts Avenue, Cambridge, MA 02139, USA}

\author{Grzegorz Nowak}
\affil{\rm Instituto de Astrof\'isica de Canarias, C/ V\'ia L\'actea s/n, E-38205 La Laguna, Spain}
\affil{\rm Departamento de Astrof\'isica, Universidad de La Laguna, E-38206 La Laguna, Spain}

\author{Enric Pall\'e}
\affil{\rm Instituto de Astrof\'isica de Canarias, C/ V\'ia L\'actea s/n, E-38205 La Laguna, Spain}
\affil{\rm Departamento de Astrof\'isica, Universidad de La Laguna, E-38206 La Laguna, Spain}

\author{Martin P\"atzold}
\affil{\rm Rheinisches Institut f\"ur Umweltforschung an der Universit\"at zu K\"oln, Aachener Strasse 209, D-50931 K\"oln, Germany}

\author{Jorge Prieto-Arranz}
\affil{\rm Instituto de Astrof\'isica de Canarias, C/ V\'ia L\'actea s/n, E-38205 La Laguna, Spain}
\affil{\rm Departamento de Astrof\'isica, Universidad de La Laguna, E-38206 La Laguna, Spain}

\author{Heike Rauer}
\affil{\rm Institute of Planetary Research, German Aerospace Center, Rutherfordstrasse 2, D-12489 Berlin, Germany}
\affil{\rm Center for Astronomy and Astrophysics, TU Berlin, Hardenbergstr. 36, 10623 Berlin, Germany}
\affil{\rm Institute of Geological Sciences, Freie Universit\"at Berlin, Malteserstr. 74-100, 12249 Berlin, Germany}

\author{Seth Redfield}
\affil{\rm Astronomy Department and Van Vleck Observatory, Wesleyan University, Middletown, CT 06459, USA}

\author{Ignasi Ribas}
\affil{\rm Institute of Space Sciences (ICE, CSIC), Campus UAB, C/Can Magrans, s/n, 08193 Bellaterra, Spain}
\affil{\rm Institut d'Estudis Espacials de Catalunya (IEEC), Barcelona, Spain}

\author{Alexis M. S. Smith}
\affil{\rm Institute of Planetary Research, German Aerospace Center, Rutherfordstrasse 2, D-12489 Berlin, Germany}

\author{Vincent Van Eylen}
\affil{\rm Department of Astrophysical Sciences, Princeton University, 4 Ivy Lane, Princeton, NJ 08544, USA}

\author[0000-0002-1623-5352]{Petr Kab\'ath}
\affil{\rm Astronomical Institute, Czech Academy of Sciences, Fri{\v c}ova 298, 251 65, Ond\v{r}ejov, Czech Republic}

\begin{abstract}
\noindent We report the discovery of an intermediate-mass transiting brown dwarf, TOI-503b, from the TESS mission. TOI-503b is the first brown dwarf discovered by TESS, and it has a circular orbit around a metallic-line A-type star with a period of $P=3.6772 \pm 0.0001$ days. The light curve from TESS indicates that TOI-503b transits its host star in a grazing manner, which limits the precision with which we measure the brown dwarf's radius ($R_b = 1.34^{+0.26}_{-0.15}\rj$). We obtained high-resolution spectroscopic observations with the FIES, Ond\v{r}ejov, PARAS, Tautenburg, and TRES spectrographs and measured the mass of TOI-503b to be $M_b = 53.7 \pm 1.2 \mj$. The host star has a mass of $M_\star = 1.80 \pm 0.06 \msun$, a radius of $R_\star = 1.70 \pm 0.05 \rsun$, an effective temperature of $T_{\rm eff} = 7650 \pm 160$K, and a relatively high metallicity of $0.61\pm 0.07$ dex. We used stellar isochrones to derive the age of the system to be $\sim$180 Myr, which places its age between that of RIK 72b (a $\sim$10 Myr old brown dwarf in the Upper Scorpius stellar association) and AD 3116b (a $\sim$600 Myr old brown dwarf in the Praesepe cluster). Given the difficulty in measuring the tidal interactions between brown dwarfs and their host stars, we cannot precisely say whether this brown dwarf formed in-situ or has had its orbit circularized by its host star over the relatively short age of the system. Instead, we offer an examination of plausible values for the tidal quality factor for the star and brown dwarf. TOI-503b joins a growing number of known short-period, intermediate-mass brown dwarfs orbiting main sequence stars, and is the second such brown dwarf known to transit an A star, after HATS-70b. With the growth in the population in this regime, the driest region in the brown dwarf desert ($35-55 M_J \sin{i}$) is reforesting.
\end{abstract}


\section{Introduction} 
\label{sec:intro}
Brown dwarfs (BDs) are loosely defined as the objects that separate giant planets from low-mass stars. This definition is based on the mass of BDs, which ranges from $11-16\mj$ (the approximate mass at which deuterium fusion can be sustained) to $75-80\mj$ (the approximate mass to sustain hydrogen fusion) and yet, some of the most recent BD discoveries seem to blur these boundaries \citep{diaz14, zhou19}. The uncertainties in boundaries are caused by dependence on exact chemical composition of objects near these mass ranges \citep{Baraffe2002, Spiegel2011}. One particular feature of the BD population is the apparent low occurrence rate of BDs in close orbits (i.e. within 3~AU) to stars in comparison to giant planets and stars. The apparent lack of short-period BDs is the so-called \emph{brown dwarf desert\/} \citep[e.g.,][]{Grether2006, Sahlmann11}. Although the population of BDs in this region has slowly grown in recent years \citep{csizmadia16}, the gap remains significant. As every desert has a driest part, the existence of a driest part of the brown dwarf desert has been argued to be the mass range between $35 \leq M_J \sin{i} \leq 55$ for orbital periods under 100 days \citep{Ma2014}. Some authors use the existence of this gap to motivate the idea that there are two separate BD populations that result from two different BD formation processes. In this case, the two processes are formation via core accretion in a protoplanetary disk (the way giant planets are thought to form) and formation by gravitational instability, which is how stars are thought to typically form. 

For core accretion to take place, an object must form in specific conditions with a sufficient gas mass budget in order for a protoplanetary core to grow sufficiently massive enough to become a giant planet or BD. This growth can be efficient at scales greater than 0.5AU, and depending on initial conditions, the giant planet or BD may then migrate inward \citep{Coleman:2017}. If significant migration occurs, then the object did not form in-situ (very near to or at its current orbit). On the other hand, if an object is on the order of $40\mj$ or more, then core accretion would not have been efficient enough to grow a protoplanetary core to that mass \citep{Mordasini:2012}. In this case, the massive object may have formed  through disk fragmentation or instabilities as low-mass stars form, making in-situ formation at a close-in orbit to the host star a more viable option than core accretion.

Regarding the current transiting BD population, we see this aforementioned gap centered at a mass of $42.5\mj$ \citep{Ma2014}. Different studies suggest suppressing the distinction between BDs and the coolest M stars given their similarities \citep{Whitworth18}, while others suggest that BDs and giant planets form a continuum based on their mass-density relation \citep{hatzes15, Persson19}, which, in turn, implies that the range of giant planets spans 0.3-60$\mj$ or 0.3-73$\mj$, respectively. Given this variety of interpretations of what separates giant planets, BDs, and low-mass stars, each new, well-characterized BD system, especially the ones that reside in the driest part of the brown dwarf desert, will be important to understanding this population as a whole.

We search for transiting BDs in particular because of the extra information that is obtained from a transiting object. In many cases, given the reliability of our stellar models, we may precisely (i.e. on the order of a few percent) measure the radius and mass of a transiting or eclipsing companion. These two properties are fundamental to an object's physical behavior and evolution. This value is enhanced for transiting BDs given that they are so uncommon and that the substellar evolutionary models that aspire to describe these objects stand to be more rigorously tested with a larger sample that has well-characterized masses and radii. With only a minimum mass provided by a radial velocity (RV) orbit, we cannot verify if a companion is truly a BD or something more massive, like a star. With only a radius that is derived from stellar models and a light curve, we cannot determine if the stellar companion is a giant planet, a BD, a low-mass star, some form of stellar activity, or a false-positive. Only with RVs and photometry combined may we identify BDs and test the mass-radius predictions of substellar evolutionary models.

This is where space-based photometric survey missions are particularly useful and often one of the best options for characterization of short-period transiting BDs.  
This was the case for the CoRoT mission \citep{Rouan1998} and the \textit{Kepler/K2} missions \citep{Borucki2010}, which made enormous contribution to exoplanetary science. So far, we are seeing a similar impact from the Transiting Exoplanet Survey Satellite (TESS\footnote{\url{https://heasarc.gsfc.nasa.gov/docs/tess/}}) mission \citep{Ricker2015} and we expect this impact to grow not only in the realm of small exoplanets, but for the transiting BD population as well. One aspect of TESS that distinguishes it from CoRoT and \textit{Kepler/K2} is the number of bright stars it will observe. This makes potential BD host stars more accessible to spectroscopic facilities that may be used in coordination with TESS as well as the {\it Gaia \/} mission (for precise parallaxes) to detect and characterize BDs. The endeavor to discover more BDs is aided further by the relatively deep transit depths of BDs around typical main sequence stars and the relatively large semi-amplitude signals relevant to RV follow-up. In total, there are more than 2,000 known BDs \citep[e.g.,][]{Skrzypek2016}, with approximately 400 of these in bound systems. Of these, only 21 transit their host stars (with an additional 2 in a BD binary, \cite{2M0535}), which makes a nearly all-sky transit survey mission like TESS an important tool in expanding and exploring the transiting BD population.

In this paper, we report the discovery of TOI-503b, the first BD known to orbit a metallic-lined A star (Am star). We find that the age of 180 Myr for TOI-503 and the circular orbit of TOI-503b are only consistent with the circularization timescale of the system for certain values of the tidal quality factor for the star and BD. However, we cannot conclusively determine which of these values best describe the system given the general uncertainty of the tidal evolution models used. This work is the result of a collaboration between the KESPRINT consortium \citep[e.g.,][]{Hjorth2019, Korth2019, Livingston2019, Palle2019, Gandolfi2019, Persson19}, PARAS-PRL India \citep{Chakraborty14}, and the Harvard-Smithsonian Center for Astrophysics. We describe the observations in Section \ref{sec:observations}, the data analysis in Section \ref{sec:analysis} and provide a final discussion in Section \ref{sec:conclusion}.

\section{Observations}\label{sec:observations}

\subsection{TESS light curves}
\begin{figure*}[!ht]
\centering
\includegraphics[width=1.0\textwidth,height=0.4\textwidth]{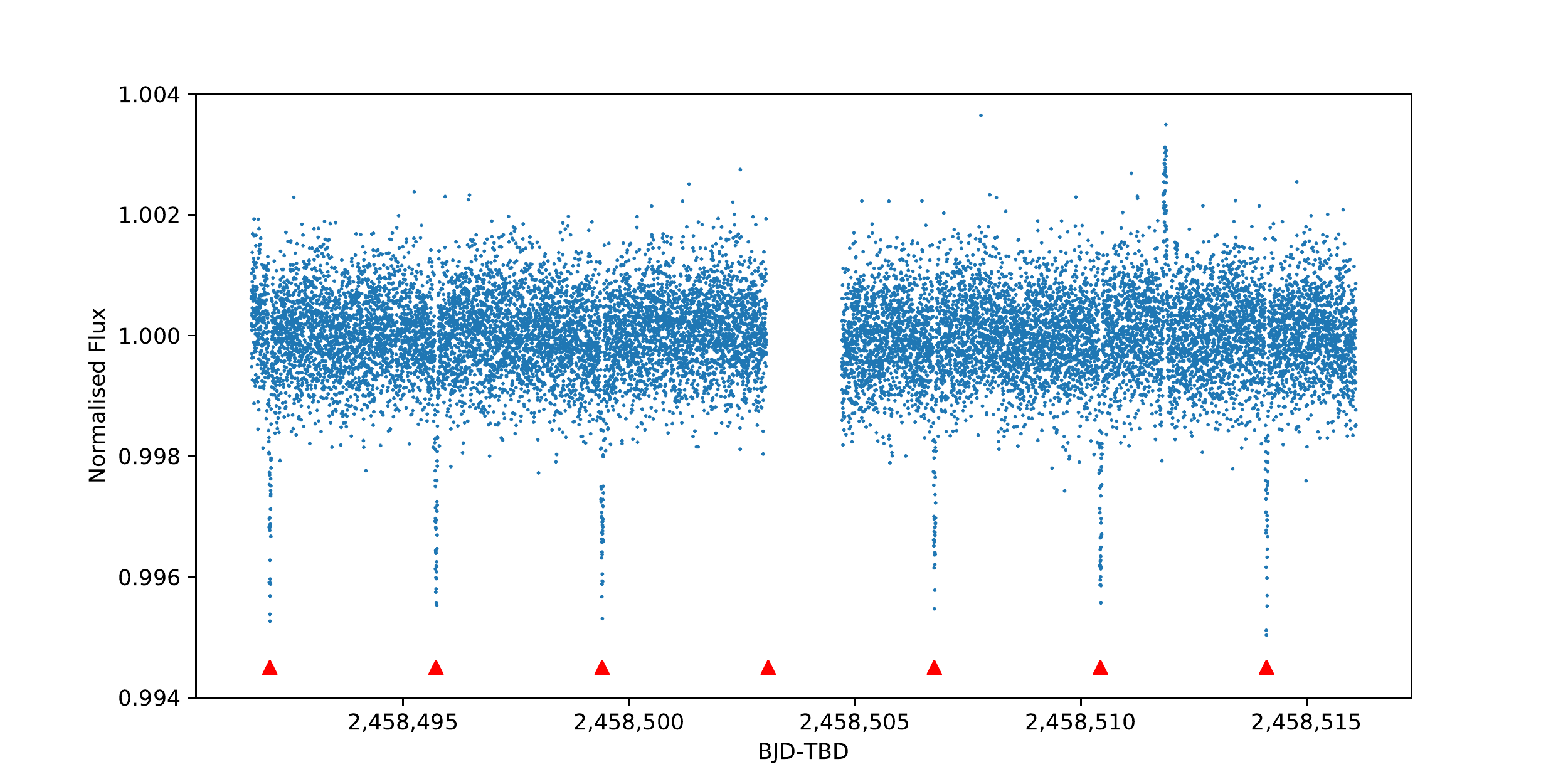}
\caption{The normalized light curve of TOI-503 observed by TESS is plotted in blue with red triangles denoting the time of each transit. Six transits can be seen spaced every $\sim$3.7 days with a depth of $\sim$4500 ppm. The bump in the lightcurve around BJD 2458511.86, is a symmetric feature from background with the duration of roughly 3.5 hrs. When excluding the star in an aperture, the bump is still visible, confirming its origin is not TOI-503.
} \label{fig:tess_lc}
\end{figure*}

TESS monitored TOI-503 at a two-minute cadence from January 8 to February 1, 2019 ($\sim$24.5 days). There is a gap of 1.7 days during this time due to the transfer of data from the spacecraft. The TESS Input Catalog (TIC) ID of the source is 186812530 \citep{stassun18} and it was observed in CCD 3 of camera 1 in Sector 7. TOI-503 will not be observed in any upcoming sectors of the primary TESS mission. We use the publicly available Pre-search Data Conditioning Simple Aperture Photometry \citep[PDCSAP;][]{stumpe2014_pdc, smith2012_pdc} light curves at Mikulski Archive for Space Telescopes (MAST)\footnote{\url{https:// mast.stsci.edu/portal/Mashup/Clients/Mast/Portal.html}} that are provided by the TESS Science Processing Operations Center (SPOC). The PDCSAP light curves have the systematics of the spacecraft removed. The SPOC pipeline \citep{Jenkins16} was used to extract the light curve and associated uncertainties from the original scientific data. We normalize this light curve by dividing it by the median-smoothed flux, which can be seen in Figure \ref{fig:tess_lc}. A total of 6 transits spaced at a period of $\sim$3.7 days are visible with depths of $\sim$4500 ppm. The TESS data validation reports \citep{Jenkins16} identify TOI-503 as the host of a planet candidate with an estimated radius of $1.13\pm0.28 \rj$ by fitting the TESS light curve and using host star parameters from \cite{stassun19}. The basic parameters of the star are listed in Table \ref{tab:basicpar}.

\begin{deluxetable}{lccc}
 \tabletypesize{\footnotesize}
 \tablewidth{0pt}
 \tablecaption{Basic parameters for TOI-503  \label{tab:basicpar}}
 \tablehead{
 \colhead{Parameter} & \colhead{Description} & \colhead{Value} & \colhead{Source}}
 \startdata
 $\alpha_{\rm J2000}$\dotfill & Right Ascension (RA)\dotfill & 08 17 16.89 & 1\\
 $\delta_{\rm J2000}$\dotfill & Declination (Dec)\dotfill & 12 36 04.76 & 1\\
 \smallskip\\
 $T$\dotfill & TESS $T$ mag\dotfill & $9.187 \pm 0.018$ & 2\\
 $G$\dotfill & Gaia $G$ mag\dotfill & $9.350 \pm 0.002$ & 1\\
 $B_T$\dotfill & Tycho $B_T$ mag\dotfill & $9.703 \pm 0.026$ & 3\\
 $V_T$\dotfill & Tycho $V_T$ mag\dotfill & $9.428 \pm 0.024$ & 3\\
 $J$\dotfill & 2MASS $J$ mag\dotfill & $8.945 \pm 0.023$ & 4\\
 $H$\dotfill & 2MASS $H$ mag\dotfill & $8.935 \pm 0.017$ & 4\\
 $K_S$\dotfill & 2MASS $K_S$ mag\dotfill & $8.895 \pm 0.016$ & 4\\
 WISE1\dotfill & WISE1 mag\dotfill & $8.868 \pm 0.023$ & 5\\
 WISE2\dotfill & WISE2 mag\dotfill & $8.885 \pm 0.020$ & 5\\
 WISE3\dotfill & WISE3 mag\dotfill & $8.888 \pm 0.029$ & 5\\
 WISE4\dotfill & WISE4 mag\dotfill & $8.558 \pm 0.020$ & 5\\
 \smallskip\\
 $\mu_{\rm \alpha}$\dotfill & PM in RA (mas/yr)\dotfill & $-9.336 \pm 0.095$ & 1\\
 $\mu_{\rm \delta}$\dotfill & PM in DEC (mas/yr)\dotfill & $-9.945 \pm 0.053$ & 1\\
 $\pi$\dotfill & Parallax (mas)\dotfill & $3.887 \pm 0.059$ & 1\\
 RV\dotfill & Systemic RV (km/s)\dotfill & $29.469 \pm 0.013$ & 6\\
 \smallskip\\
 \hline
 \hline
 \multicolumn{2}{l}{Other identifiers:} & \smallskip\\
 & TIC 186812530\\
 & TYC 802-751-1\\
 & 2MASS J08171689-1236049\\
 & Gaia DR2 650254479499119232\\
 \enddata
 \tablecomments{References: 1 - \citet{Lindegren2018}, 2 - \citet{stassun18},\\ 3 - \citet{Hog2000}, 4 - \citet{Cutri2003}, 5 - \citet{Cutri2013},\\ 6 - this work}
\end{deluxetable}

\subsection{Ground-based light curves}
As part of the TESS Follow-up Observation Program (TFOP) additional ground-based photometry was carried out by the Sinistro camera on the Las Cumbres Observatory (LCO), Siding Spring Observatory (SSO)\\ 1.0-m on March 19, 2019, the Santa Barbara Instrument Group (SBIG) camera on the LCO 0.4-m on March 19, 2019, the Chilean Hungarian Automated Telescope (CHAT) 0.7-m telescope on March 22, 2019, and the \textit{KeplerCam} instrument on the Fred Lawrence Whipple Observatory (FLWO) 1.2-m telescope on April 25, 2019. The LCO-SSO observations were taken in the Y-band and confirmed that there are no nearby or background eclipsing binaries within $2\farcm5$ that were blended in the aperture of camera 1 from TESS. The transit was not detected by LCO-SSO due to the insufficient amount of out-of-transit baseline flux. The observations with SBIG show a clear ingress, but do not extend long enough to show the egress of the transit due to the target star reaching a high airmass. A full on-time transit was detected by CHAT in the $i\,\acute{}$ band as well as the \textit{KeplerCam} instrument in the $z$ band. By independently fitting just the \textit{KeplerCam} light curve using {\tt AstroImageJ} \citep{Collins:2017}, we find that the modeled transit center time is consistent with the time predicted by the public TESS ephemeris within 1-$\sigma$ uncertainty. We decide against incorporating any ground-based follow-up in our joint analysis due to the shallow nature of this transit and the low transit depth signal-to-noise ratio. 

\subsection{Contamination from nearby sources}\label{sec:ao_image}
\begin{figure}[!ht]
\centering
\includegraphics[width=0.46\textwidth, trim= {0.0cm 0.0cm 0.0cm 0.0cm}]{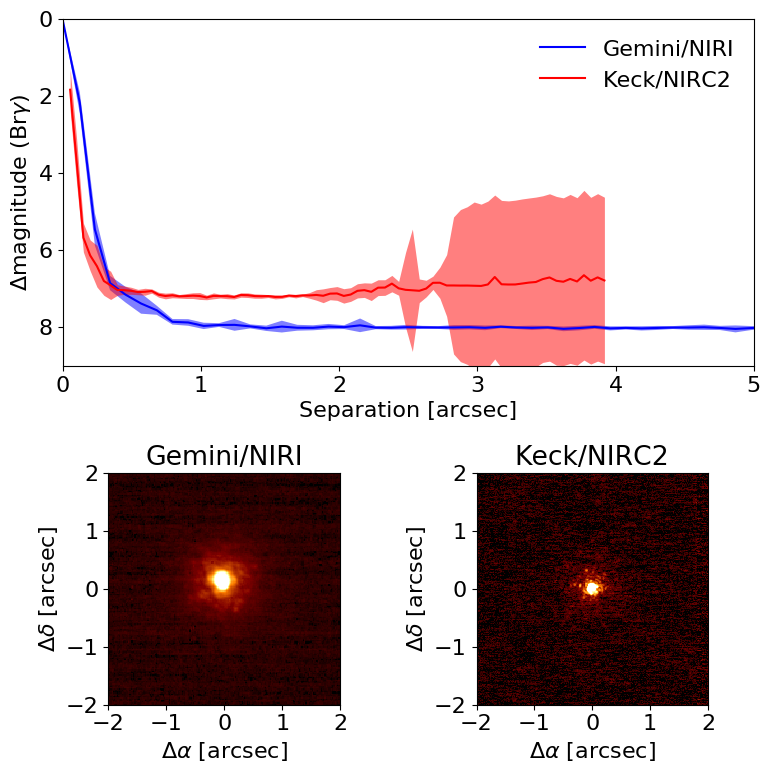}
\caption{Sensitivity curve as a function of angular separation for TOI-503 from Gemini/NIRI and Keck/NIRC2. The inset shows the image of the target star from each instrument.
} \label{fig:ao_images}
\end{figure}

The TFOP was also responsible for observations of TOI-503 with Gemini/NIRI on March 22, 2019 and again with Keck/NIRC2 \citep{wizinowich2000} on April 7, 2019 (Figure \ref{fig:ao_images}). In each case, observations were taken in NGS mode in the Br-$\gamma$ filter with the target as the guide star. Images were dithered, such that a sky background could be constructed, with a square dithering pattern for the NIRI data and a 3-point pattern for the NIRC2 data to avoid the known noisy fourth quadrant. For each instrument we used the same basic reduction procedure: images were flat-fielded and sky-subtracted, and the dithered frames aligned and co-added.

Sensitivity was determined by injecting simulated sources azimuthally around the primary target, at separations of integer multiples of the central source's full width at half maximum \citep{furlan2017}. The brightness of each injected source was scaled until standard aperture photometry detected it with 5-$\sigma$ significance. The resulting brightness of the injected sources relative to the target set the contrast limits at that injection location. The final $5\sigma $ limit at each separation was determined from the average of all of the determined limits at that separation and the uncertainty on the limit was set by the rms dispersion of the azimuthal slices at a given radial distance. No nearby contaminating sources are identified in either image, and at $1\arcsec$ we reach contrasts of $\Delta$mag=8.0mag in the NIRI data and $\Delta$mag=7.2mag in the NIRC2 data.

\subsection{KESPRINT spectra}
We obtained a total of 50 spectra of TOI-503 between March 18, 2019 and April 17, 2019 using KESPRINT observing time on the 2-m Perek telescope at the Ond\v{r}ejov Observatory, the 2-m Alfred Jensch telescope at Tautenburg, and the 2.56-m Nordic Optical Telescope (NOT) at the Roque de Los Muchachos Observatory. Using the central Europe monitoring network with telescopes in Ond\v{r}ejov and Tautenburg for simultaneous observations has the advantage to allow a better coverage of observing data. Furthermore, both telescopes are capable of long term monitoring of interesting objects \citep{Kabath19}. For these reasons, such observations are often performed \citep{Skarka2019,Kabathetal2019,Sabotta2019}. RVs from all used telescopes beyond the KESPRINT are reported in Table \ref{tab:toi503_rvs}. 

\subsection{Ond\v{r}ejov spectra}
We collected a set of 14 spectra using the Ond\v{r}ejov Echelle Spectrograph, which has a spectral resolving power R\,$\approx$\,44\,000 over the wavelength range of 370nm to 850nm \citep{Kabath19}. All spectra have an exposure time of 3600\,s resulting in a signal-to-noise (S/N) per pixel at 550nm varying between 16--22, depending on the observing conditions and the airmass. We use the standard IRAF 2.16 routines \citep{Tody1993} to process the spectra, which were corrected for bias, flat field and cosmic rays. The spectrum with the highest S/N was used as template for the cross-correlation done with the IRAF {\tt fxcor} routine, allowing us to remove instrumental shift by measuring the shift in telluric lines, and to measure the relative RVs. The errors are standard deviations of values from eighteen 10nm intervals that were considered.

\subsubsection{FIES spectra}
We acquired 8 spectra with the FIbre-fed {\'E}chelle Spectrograph \citep[FIES;][]{1999anot.conf...71F,2014AN....335...41T} mounted at the 2.56-m Nordic Optical Telescope (NOT) of Roque de los Muchachos Observatory (La Palma, Spain). FIES has a resolving power of R\,$\approx$\,47\,000. The observations were carried out between March 21 and April 15, 2019 UT, as part of the observing programs 58-024 and 59-210. The exposure time was set to 1500--2100 s -- depending on sky and seeing conditions --, leading to a S/N ratio per pixel of $\sim$70-100 at 5500\,\AA. We followed the observing strategy described in \cite{2010ApJ...720.1118B} and \cite{2015A&A...576A..11G} and traced the RV drift of the instrument by acquiring long-exposed ThAr spectra ($T_\mathrm{exp}$\,$\approx$\,60\,s) immediately before and after each science exposure. We reduced the FIES spectra following standard IRAF and IDL routines and extracted the RV measurements via multi-order cross-correlations with the RV standard star HD\,182572 \citep{1999ASPC..185..367U} observed with the same instrument set-up as TOI-503.

\subsubsection{Tautenburg spectra and Doppler Tomography analysis}
We used the 2-m Alfred Jensch telescope of the Th\"uringer Landessternwarte Tautenburg to obtain 28 spectra of TOI-503. The telescope is equipped with an echelle spectrograph with spectral resolving power R\,$\approx$\,35\,000 with the $2\arcsec$ slit used. The spectra used for orbital analysis have an exposure time 1200\,s resulting in an S/N ratio between 23 and 27. We processed the spectra using the Tautenburg Spectroscopy pipeline \citep{Sabotta2019} built upon {\tt PyRaf} and the Cosmic Ray code by Malte Tewes based on the method by \cite{Dokkum2001}. We use cross-correlation routines from IRAF to correct spectra for the shift in telluric lines and to measure the relative RVs. There are 17 spectra from the 28 which have an exposure time of 600\,s and that were taken in an attempt to extract a Doppler tomography (DT) \citep[e.g.,][]{hatzes1998, albrcht2007, collier2010} signal during the transit night of April 17, 2019. These are not used for the RV measurements to avoid the signal created by the BD blocking light from the host star, which creates an additional Doppler shift that is based on the orbital alignment and rotation rate of the star and not the orbital motion of the BD.

The DT technique reveals the distortion of the stellar line profiles when a planet or BD blocks part of the stellar photosphere during a transit. This distortion is a tiny bump in the stellar absorption profile, scaled down in width according to the BD-to-star radius ratio. Additionally, the area of that bump corresponds to the BD-to-stellar disks area ratio. As the BD moves across the stellar disk, the bump produces a trace in the time series of line profiles, which reveals the spin-orbit alignment between the star and BD orbit. For this analysis, we first created a reference stellar absorption spectrum consisting of delta functions at the wavelength positions of the observed stellar absorption lines. Their positions and strengths were determined by fitting each stellar absorption line in the observed spectrum with the rotational profile of TOI-503 ($v\sin{i} = 26{\rm kms^{-1}}$). A total of 410 stellar absorption lines were identified in the wavelength range from 455.8 to 674.6nm. We excluded those wavelength regions from our analysis which exhibited telluric lines, the Hydrogen Balmer absorption lines and the Na II doublet around 589nm.

\begin{figure}[!ht]
\centering
\includegraphics[width=0.47\textwidth, trim= {0.0cm 0.0cm 0.0cm 0.0cm}]{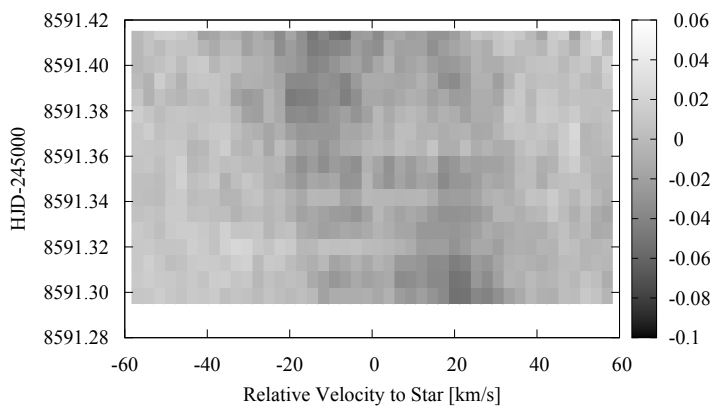}
\caption{Doppler tomography using Tautenburg in-transit spectra.} \label{fig:dt_tautenburg}
\end{figure}

By employing a least-squares deconvolution similar to what is shown in \cite{collier2002} of the observed spectra with the reference spectrum, we summed up the 410 stellar absorption lines in each spectrum into one high S/N mean line profile. The resulting line profiles were scaled so that their height was one, and were interpolated onto a velocity grid of 2.65 km/s increments, corresponding to the velocity range of one spectral pixel at 550nm. We then summed up all the mean line profiles collected the nights before the transit and subtracted the resulting profile from the in-transit ones. Figure \ref{fig:dt_tautenburg} shows the residuals of the line profiles and shows that we are unable to detect a trace of the transiting planet using this method. 

\begin{deluxetable}{cccc}
\tabletypesize{\footnotesize}
\tablewidth{0pt}

 \tablecaption{Multi-order relative radial velocities of TOI-503 from Ond\v{r}ejov, FIES, Tautenburg, TRES, and PARAS. \label{tab:toi503_rvs}}

 \tablehead{
 \colhead{$\rm BJD_{\rm TDB}-2450000$} & \colhead{RV (m/s)} & \colhead{$\sigma_{\rm RV}$ (m/s)} & \colhead{Instrument}}

\startdata 
8566.651866 & -24.9 & 84.6  & TRES \\
8568.628445 & 8720.4 & 87.6 & TRES \\
8569.654306 & 1721.4 & 76.6 & TRES \\
8570.631553 & 1140.6 & 35.6 & TRES \\
8571.619256 & 8129.6 & 52.7 & TRES \\
8572.647920 & 6927.5 & 71.7 & TRES \\
8573.712391 & 30.2 & 54.1   & TRES \\
8574.660766 & 3615.3 & 65.4 & TRES \\
8575.644821 & 9286.9 & 51.2 & TRES \\
8576.674889 & 4227.1 & 59.9 & TRES \\
8577.649209 & -64.4 & 47.5  & TRES \\
8587.709792 & 4140.1 & 63.7 & TRES \\
8581.233706 & -4877.9 & 92.1  & PARAS\\
8582.207307 & 539.1 & 70.9  & PARAS\\
8582.238727 & 805.1 & 88.6  & PARAS\\
8583.212860 & 4393.9 & 85.7  & PARAS\\
8583.242544 & 4175.4 & 100.8 & PARAS\\
8584.201377 & -1800.3 & 86.5  & PARAS\\
8585.220041 & -3730.9 & 87.4  & PARAS\\
8564.408181 & 7370.8 & 250.5 & OND\v{R}EJOV\\
8564.450341 & 7663.6 & 191.3 & OND\v{R}EJOV\\
8565.411871 & 5208.5 & 292.7 & OND\v{R}EJOV\\
8565.454031 & 4899.8 & 452.4 & OND\v{R}EJOV\\
8566.414342 & -1016.0 & 405.8 & OND\v{R}EJOV\\
8572.314192 & 7360.1 & 391.9 & OND\v{R}EJOV\\
8575.276802 & 6639.4 & 291.2 & OND\v{R}EJOV\\
8575.369402 & 7511.8 & 227.1 & OND\v{R}EJOV\\
8575.419602 & 7484.1 & 305.0 & OND\v{R}EJOV\\
8578.300872 & 2048.7 & 274.6 & OND\v{R}EJOV\\
8578.343032 & 2911.4 & 199.8 & OND\v{R}EJOV\\
8578.385192 & 3016.4 & 222.3 & OND\v{R}EJOV\\
8581.353282 & -674.9 & 427.5 & OND\v{R}EJOV\\
8581.395442 & -829.4 & 409.7 & OND\v{R}EJOV\\
8559.414221 & -7002.3 & 212.0 & TAUTENBURG\\
8561.366911 & 1000.0 & 80.0 & TAUTENBURG\\
8562.458981 & -6541.5 & 200.3 & TAUTENBURG\\
8563.348001 & -5693.6 & 143.8 & TAUTENBURG\\
8567.360682 & -3341.8 & 128.5 & TAUTENBURG\\
8589.311442 & -4202.8 & 160.8 & TAUTENBURG\\
8589.326082 & -3905.7 & 160.8 & TAUTENBURG\\
8589.340732 & -4058.3 & 82.7 & TAUTENBURG\\
8590.312772 & 1848.7 & 200.9 & TAUTENBURG\\
8590.326992 & 1853.5 & 185.5 & TAUTENBURG\\
8590.341222 & 1665.8 & 273.4 & TAUTENBURG\\
8564.442588 & 33708.6 & 26.8 & FIES\\
8566.400806 & 24939.9 & 33.5 & FIES\\
8581.364308 & 24956.5 & 36.1 & FIES\\
8583.434302 & 33152.1 & 32.4 & FIES\\
8587.367559 & 31681.2 & 21.5 & FIES\\
8587.474228 & 30949.2 & 27.0 & FIES\\
8588.453630 & 24951.0 & 45.7 & FIES\\
8589.369006 & 28516.5 & 47.5 & FIES\\
\enddata
\end{deluxetable}

\subsection{TRES spectra}
We used the Tillinghast Reflector Echelle Spectrograph (TRES) on Mt. Hopkins, Arizona to obtain spectra of TOI-503 between March 23 and April 14, 2019. The spectrograph has a resolving power of R\,$\approx$\,44\,000 and covers wavelengths from 390nm to 910nm. Forty-three spectra of TOI-503 were taken with TRES with exposure times ranging from 195--300 s and S/N ranging from 35 to 59. The relative RVs that we derive from TRES spectra use multiple echelle orders from each spectrum that are cross-correlated with the highest S/N spectrum of the target star. We omit individual orders with poor S/N and manually remove obvious cosmic rays. Of these 43 spectra, 33 were taken in an attempt to extract a DT signal, but as with our analysis of the Tautenburg in-transit DT spectra, we do not find a noticeable signal.

\subsection{PARAS spectra}
We obtained 7 spectra with the PARAS spectrograph \citep{Chakraborty14} coupled with the 1.2-m telescope at Gurushikhar Observatory, Mount Abu, India between April 6 to April 11, 2019 at a resolving power of R\,$\approx$\,67\,000, in the wavelength range of 380nm to 690nm. Each night had a median seeing of around 1.5\arcsec. The exposure time for each measurement was kept at 1800 s, which resulted in a S/N of 20--25 at the blaze peak wavelength of 550nm. The spectra were extracted using a custom-designed automated pipeline written in IDL, based on the algorithms of \cite{Piskunov02}. The extracted spectra were cross-correlated with the template spectrum of an A-type star to calculate the relative RVs. Further details of the spectrograph and data analysis procedure can be found in \cite{Chakraborty14}. The uncertainties reported here are the cross-correlation function fitting errors combined with the photon noise in the same way as described in \cite{Chaturvedi2016, Chaturvedi18}.

\section{Analysis} \label{sec:analysis}

\subsection{Modeling Stellar Parameters}
We use {\tt iSpec} \citep{BlancoCuaresma, BlancoCuaresma2019} and the Stellar Parameter Classification ({\tt SPC}) software \cite{spc} to analyze the spectra of TOI-503. Then, with the spectral properties as well as an SED, light curve, and {\it Gaia \/} parallax of the star, we use {\tt EXOFASTv2} \citep{exofastv2}, and combination of {\tt PARAM 1.3} (to model the stellar parameters) \citep{daSilva06}, {\tt GeePea} (to model the light curve) \citep{Gibson12} and {\tt Systemic Console} (to model the RV curve) \citep{Meschari2009} to independently model the star and BD.

\begin{figure*}[!ht]
\centering
\includegraphics[width=1.0\textwidth,height=0.75\textwidth]{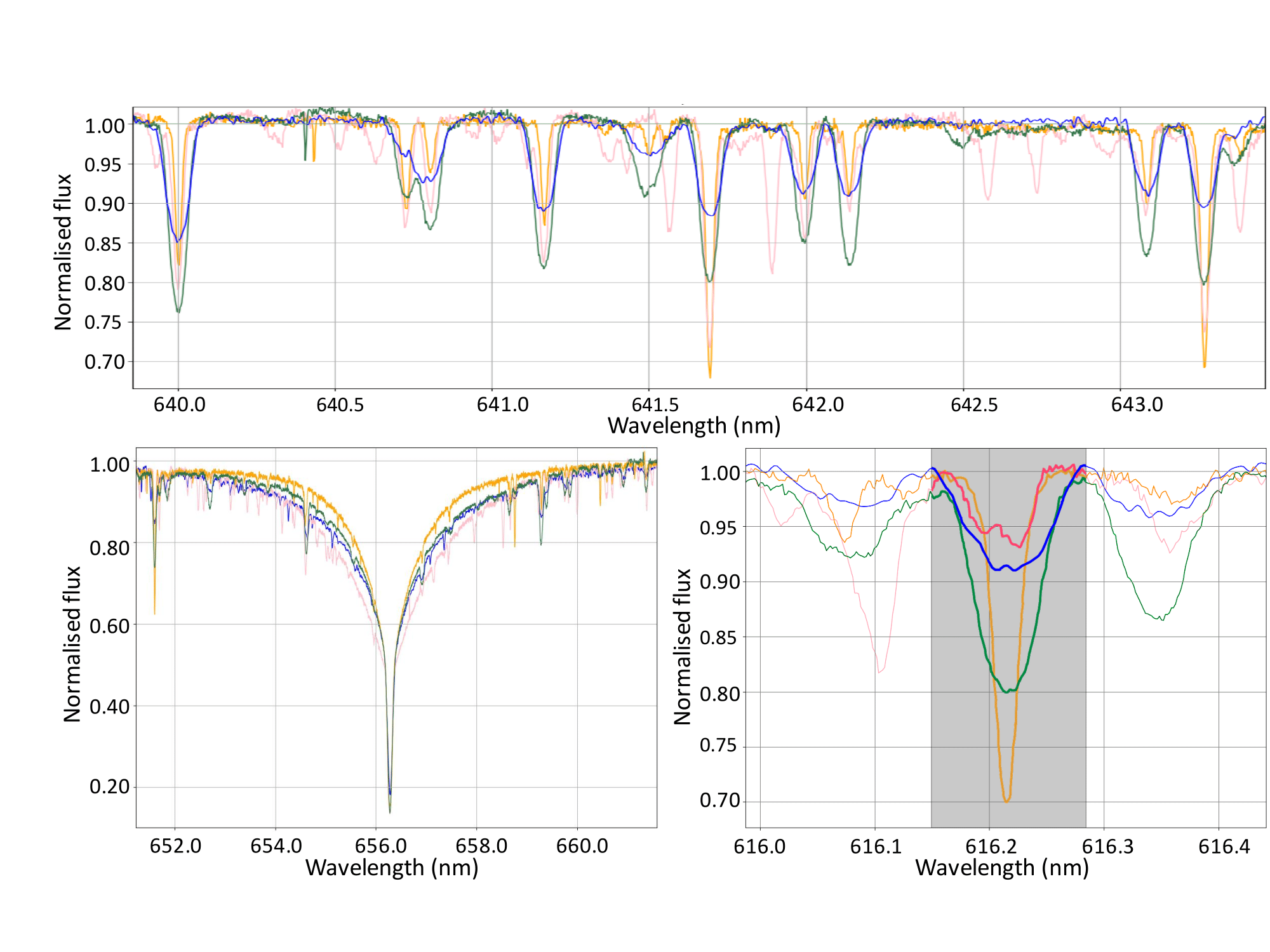}
\caption{Spectrum of TOI-503 (blue line) and three templates with similar temperatures (Am-star template -- green line, A-star template -- orange line, Ap-star template -- pink line), over-plotted for comparison. Top: iron lines region. Bottom left: the H$\alpha$ Balmer line. Bottom right: a CaI line highlighted by grey region.
} \label{fig:spectrum}
\end{figure*}

\subsubsection{{\tt iSpec} Stellar Parameters} \label{iSpec}
We use {\tt iSpec} to perform a detailed analysis of the host star from the FIES spectra. Specifically, we use the Synthe radiative transfer code \citep{Kurucz93}, the MARCS atmosphere models \citep{Gustafsson08}, and version 5 of the GES atomic line list \citep{Heiter2015} between 420 and 920nm, which includes 35 different chemical species. These are incorporated into the framework of {\tt iSpec}. We co-add all the 8 FIES spectra (after the RV shift correction) to increase the S/N and use them to determine the effective temperature $T_{\rm eff}$, metallicity $\rm [Fe/H]$, surface gravity $\log{g}$, and the projected stellar equatorial velocity $v\sin{i}$. We model the stellar parameters using the Bayesian parameter estimation code {\tt PARAM 1.3} and use the parallax measured by {\it Gaia \/} DR2 \citep[$\varpi= 3.8875 \pm 0.0591\,\rm$ mas;][]{Lindegren2018} for the distance of the star and Tycho V magnitude \citep{Hog2000}. {\tt PARAM 1.3} code estimates stellar properties using the PARSEC isochrones \citep{Bressan2012}. We calculate the value of $\log{g}$ iteratively to ensure an agreement between {\tt iSpec} and {\tt PARAM 1.3}. We determine the effective temperature by fitting the H$\alpha$ Balmer line \citep{Cayrel2011}, and the metallicity by fitting for 22 Fe I lines in the interval 597nm to 643nm. From this analysis, we find TOI-503 to be a metallic-line A star, or Am star, with a metallicity of $[{\rm Fe/H}]=0.61 \pm 0.07$. 

The formation of Am stars is generally attributed to the slowing of the stellar rotation via tidal force caused by a binary star \citep{Michaud83}. Am stars are generally slow rotators compared to typical A stars, with rotation rates below $120$ km/s \citep{Abt95}. The study by \cite{Abt73} suggests that all slowly rotating A-type main sequence stars are chemically peculiar, i.e. those with high iron abundance or unusual an depletion of key elements such as Ca. The rotation period of TOI-503 ($P_{\rm rot}$ = 3.64 days) is determined from the projected stellar equatorial velocity, the inclination derived from {\tt GeePea} and the radius of the star derived from {\tt PARAM 1.3}. The rotation period of the star is similar to the orbital period of the BD ($P_{\rm orb}$ = 3.67 days), which is indicative of synchronism. This analysis assumes the alignment between the equatorial and orbital planes of the star. However, such an assumption is not surprising for the close binary system like TOI-503. For example, the paper by \citet{Hale1994} suggests approximate alignment for solar-type binaries under the separation of 30-40 AU. The paper by \cite{hut1980} shows that for close binary systems where the tidal evolution is the primary mechanism linked with temporal changes in orbital parameters, the tidal equilibrium can be established only under assumptions of coplanarity, circularity, and synchronised rotation.  Furthermore, the similarity with the rotation period determined from the photometric light curve in section \ref{sec:rotation} also reflects approximate alignment. Such a slow rotation rate would enable the onset of radiative diffusion within the stable atmosphere, which leads to the abundance of elements observed in the spectrum, as in Am stars \citep{Michaud83}. In this context, comparing the TOI-503 spectrum with the templates of a normal A-type star, a magnetically peculiar Ap star, and an Am star from the ESO database\footnote{\url{http://www.eso.org/sci/observing/tools/uvespop/field_stars_uptonow.html}} reveals the clear similarity between the observed spectrum of TOI-503 and that of the Am stars (Figure \ref{fig:spectrum}). However, the most persuasive argument would be that the overabundance (in context of A-type stars) of the iron group elements is coupled with an underabundance of key light elements, such as Ca, Sc, or Mg, which is the characteristic sign of Am stars. The abundances we derive point exactly to this conclusion, thereby confirming the Am classification. The stellar parameters and the abundances of selected species are reported in Table \ref{tab:compare}. 

\subsubsection{Stellar Parameter Classification and {\tt EXOFASTv2} modeling}
We also use {\tt SPC} with the TRES spectra to independently (from {\tt iSpec} and FIES) derive effective temperature ($T_{\rm eff}$), metallicity ($\rm [Fe/H]$), surface gravity ($\log{g}$), and the projected stellar equatorial velocity ($v\sin{i}$) for TOI-503. We iteratively use {\tt SPC} with {\tt EXOFASTv2} \citep{exofastv2} to determine values for $T_{\rm eff}$ and [Fe/H], meaning that we use the $\log{g}$ from {\tt EXOFASTv2} as a fixed parameter in {\tt SPC} and then take the $T_{\rm eff}$ and [Fe/H] from {\tt SPC} (with the fixed $\log{g}$ from {\tt EXOFASTv2}) as starting values in a new {\tt EXOFASTv2} analysis. However, due to the upper limit of $\rm [Fe/H] \leq +0.5$ for the metallicity of the MIST isochrones \citep{mist1, mist2, mist3} that {\tt EXOFASTv2} utilizes, we rely on our measurements using {\tt iSpec} for the metallicity. With {\tt SPC}, we measure a metallicity of $[{\rm Fe/H}]=0.34 \pm 0.08$ with a fixed $\log{g} = 4.23$ from an initial {\tt EXOFASTv2} analysis. The parameters $T_{\rm eff}$, [Fe/H], and $\log{g}$ are not fixed in subsequent {\tt EXOFASTv2} analyses and only $\log{g}$ is fixed in subsequent {\tt SPC} anaylses. The [Fe/H] value is about 0.27 dex lower than our value from {\tt iSpec} most likely because {\tt SPC} explicitly measures [m/H], which is a good approximation for [Fe/H] assuming a Solar-like composition and chemical proportions (not the case for TOI-503). We use {\tt SPC} on a co-added spectrum with the RV shifts corrected for. We do not co-add any spectrum with S/N$<$15. With {\tt SPC}, we use the $503\rm $-$532$nm wavelength range (centered on the Mg b triplet) on a single co-added TRES spectrum.

\begin{figure}[!ht]
\centering
\includegraphics[width=0.44\textwidth, trim= {0.0cm 0.0cm 0.0cm 0.0cm}]{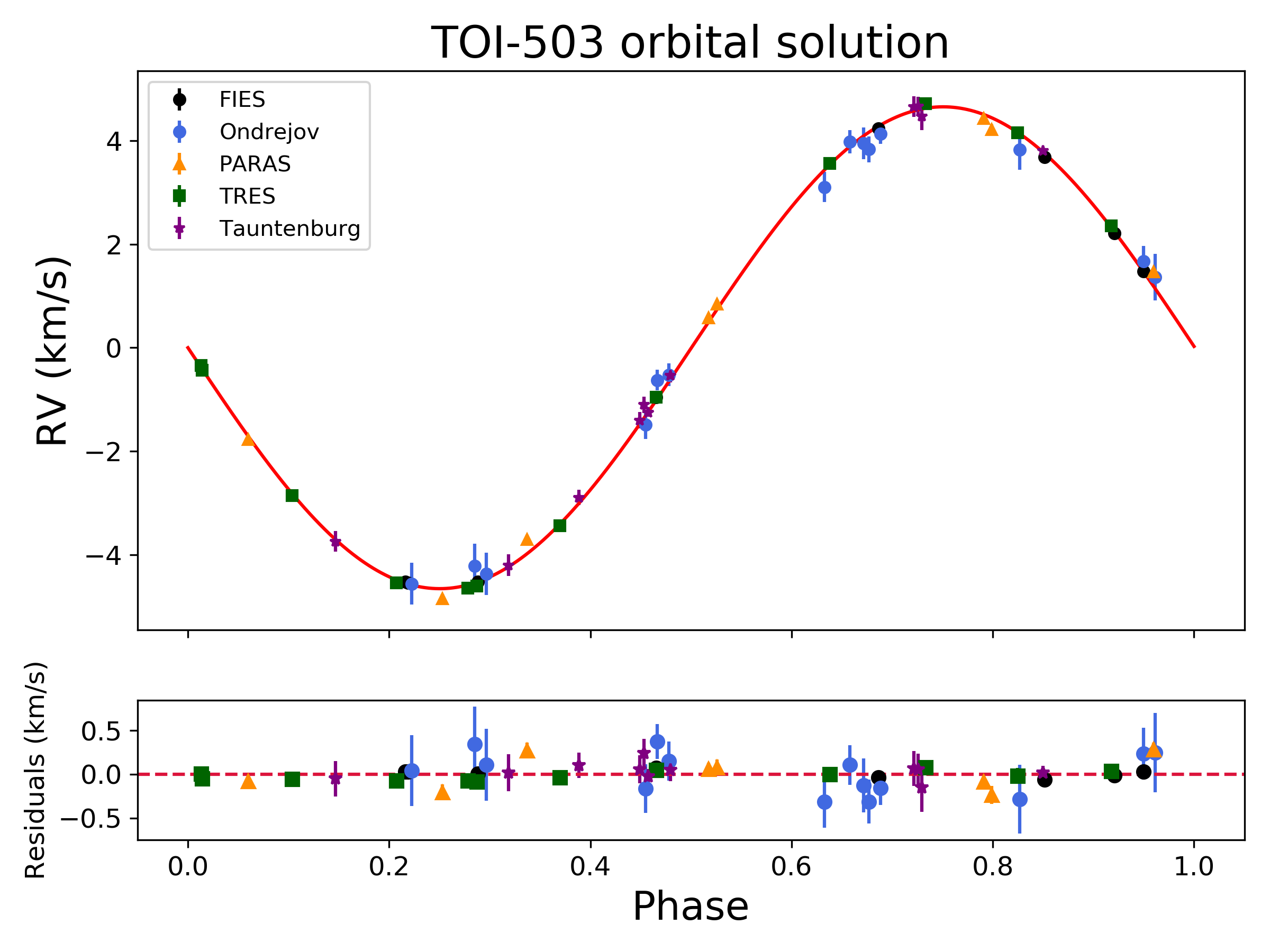}
\caption{Orbital solution for TOI-503 showing the {\tt EXOFASTv2} RV model in red. This orbital solution is jointly derived by simultaneously fitting all RVs from the different contributing spectrographs and the normalized PDCSAP TESS light curve.} \label{fig:orbital_solution}
\end{figure}

We derive the mass and radius of the BD using {\tt EXOFASTv2}, which uses the Monte Carlo-Markov Chain (MCMC) method. For each MCMC fit, we use N=36 (N = 2$\times n_{\rm parameters}$) walkers, or chains, and run for 50,000 steps, or links. We modeled the host star mass and radius using the MIST isochrones, which are integrated into the framework of {\tt EXOFASTv2}. Figure \ref{fig:orbital_solution} shows the orbital solution we derive with {\tt EXOFASTv2} with a joint fit of the RV and transit data. Our transit solution from this same joint fit agrees with that shown via the {\tt GeePea} analysis (Figure \ref{fig:LC_fittting}). We account for interstellar extinction, $A_V$, using the Galactic dust and reddening extinction tool from IRAS and COBE/DIRBE \footnote{Galactic dust and reddening extinction tool: \url{https://irsa.ipac.caltech.edu/applications/DUST/}} and take this value of $A_V=0.0791$ as an upper limit for our priors in {\tt EXOFASTv2}. We also use the parallax of TOI-503 as measured by {\it Gaia \/} DR2 and the {\tt SPC} results for $T_{\rm eff}$ and metallicity ($\rm [Fe/H]=0.34$) as starting points for our priors. The full list of free parameters we specify for each object is: period $P$, time of conjunction ($T_C$ in BJD), host star effective temperature $T_{\rm eff}$, host star metallicity [Fe/H], RV semi-amplitude $K$, RV relative offset to the systemic velocity $\gamma_{rel}$, interstellar extinction $A_V$, parallax, orbital inclination $i$, and $R_B/R_\star$. We initially allow the eccentricity $e$ to be a free parameter and find it to be close to zero at $e\approx 0.007 \pm 0.003$ ($ e\cos{\omega}=-0.0058\pm 0.0032$, $ e\sin{\omega}=-0.0004\pm 0.0037$). In order to avoid the Lucy-Sweeney bias \citep{lucysweeny}, we fix the eccentricity to zero in all subsequent analyses. The derived $T_{\rm eff}$ from {\tt EXOFASTv2} agrees well with the spectroscopic $T_{\rm eff}$ from {\tt SPC}. We impose Gaussian priors on these free parameters in {\tt EXOFASTv2}. The median value with 1-$\sigma$ uncertainties of the MCMC chains for each parameter is reported in Table \ref{tab:exofast_table}. The parameters derived from {\tt EXOFASTv2} are consistent with those derived from our other independent analyses.

\subsubsection{{\tt Systemic Console} and {\tt GeePea} modeling}\label{sec:geepea}

Besides {\tt EXOFASTv2}, we also use the {\tt Systemic Console} package to model the orbital solution additionally. We consider data from each spectrograph with corresponding velocity offsets to be free parameters allowing us to fit all datasets simultaneously. By synergy of the Lomb-Scargle (LS) periodogram and Levenberg-Marquardt minimization method, we found the best solution providing the starting values for an MCMC analysis with four chains with 1,000 walkers of 50,000 iterations. As the SPOC LC gives us a better estimate of the orbital period, we fix this value in all analyses. Similarly to {\tt EXOFASTv2} modeling, we also fix the eccentricity to zero to avoid the Lucy-Sweeney bias. The median values of the main parameters with 1-$\sigma$ uncertainties of the MCMC chains, together with values from {\tt EXOFASTv2} modeling, are reported in Table \ref{tab:compare}. The parameters derived from the {\tt Systemic Console} are consistent with those derived from our other independent analyses.

TOI-503 has a V-shaped, grazing transit. In general, V-shaped eclipses are often considered false positives caused by binary stars with similar radii, stellar grazing eclipses, or a blended eclipsing binary, such as a background binary or one bound to the target star in a triple system. In this case, we ruled out the possibility of a false positive scenario with a combination of follow-up RVs to determine the mass of the companion and high-resolution imaging to rule out a blend. This implies a rather low inclination in context of BDs that we measure to be roughly $i = 82.25^\circ \pm 0.41$ or an impact parameter of $b=0.974^{+0.022}_{-0.015}$. There are slightly more than 10 similar systems known \citep{Alsubai2018}, but only one that includes a BD \citep{Czizmadia2015}. The analysis of grazing eclipses is rather challenging and often degenerate between the radius of the transiting object and its impact parameter $b$. 

\begin{figure}[!ht]
\centering
\includegraphics[width=0.44\textwidth, trim= {0.0cm 0.0cm 0.0cm 0.0cm}]{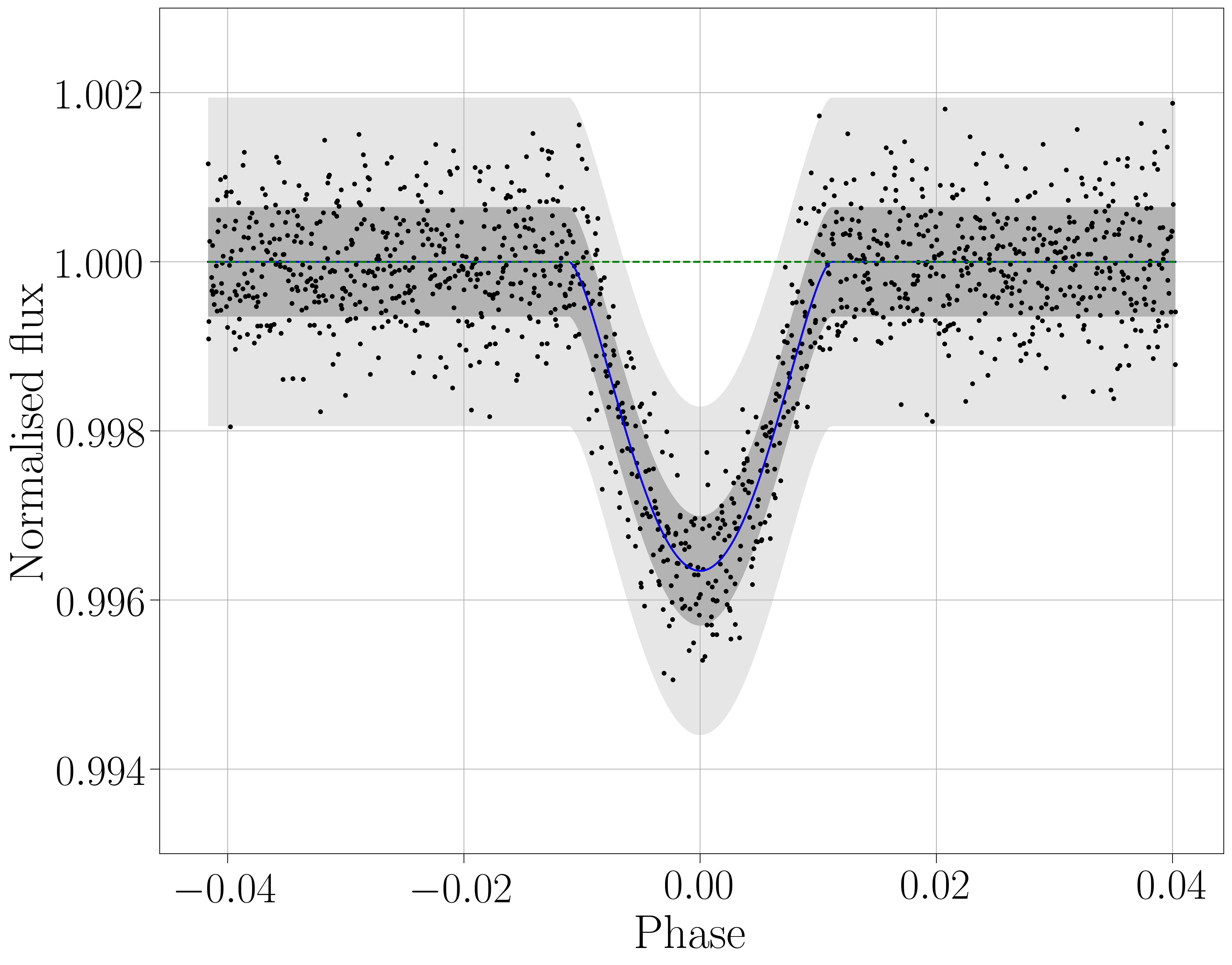}
\caption{The transit light curve of the TOI-503, fitted with the GP model described in Sect. \ref{sec:geepea}. The blue line represents the best fitting transit light curve and the green line shows the model without the transit function. The dark and light grey regions represent the 1-$\sigma$ and 3-$\sigma$ prediction of the GP model.} \label{fig:LC_fittting}
\end{figure}

\begin{figure*}[!ht]
\centering
\includegraphics[width=1.0\textwidth,height=1.0\textwidth]{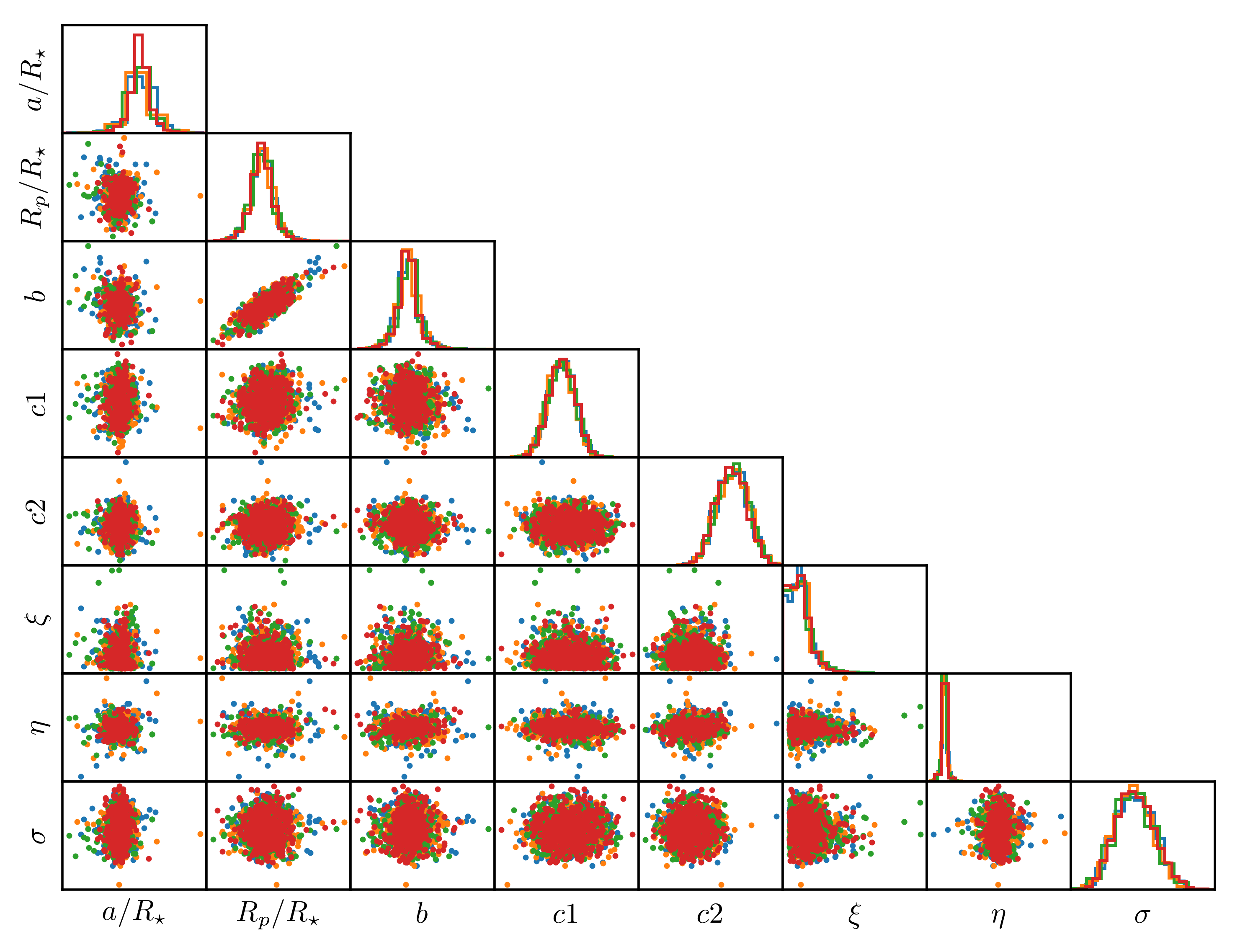}
\caption{The correlations between the free parameters of the LC model from the MCMC analysis using the {\tt GeePea} code. At the end of each row is shown the derived posterior probability distribution. We use the quadratic limb-darkening law with the coefficients $c_{1}$ and $c_{2}$. $\xi$ (output scale describing the GP’s variance), $\eta$ (length scale that determines smoothness of the function), $\sigma$ (Poisson noise) are the parameters of the noise model. The four different colours represent samples from the independent MCMC chains.
} \label{fig:correlation}
\end{figure*}

We fit the light curve using the {\tt GeePea} code, which is based on Gaussian Processes (GPs) and described by \cite{Gibson12}. We use square exponential kernel function and assume uniform, uninformative priors for all the parameters of the transit and noise model with the additional restrictions for the limb darkening coefficients chosen so as to ensure a positive surface brightness, and for all hyper-parameters of the noise model chosen such as to ensure their values are positive. Since limb darkening, radius ratio, and impact parameter are degenerate here due to the grazing transit geometry, the fitted values for the radius ratio and impact parameter dominate the limb darkening measurement. Considering this, we set Gaussian priors on the limb darkening coefficients obtained from the tables of \cite{Claret2017}. The fit of the light curve is presented in Figure \ref{fig:LC_fittting}. We use the MCMC method with four chains with 1,000 walkers of 40,000 iterations to find out the uncertainties for the each parameter of the transit and noise model. Plots of posterior distributions and correlation plots are presented in Figure \ref{fig:correlation}. The determined values for the parameters are summarized in Table \ref{tab:compare} and are found independently from those found using {\tt EXOFASTv2}.

We also use other analysis tools to perform independent analyses of the RV and transit data and derive the stellar and the BD parameters of TOI-503: {\tt PYANETI} \citep{Barrag2019} and {\tt MISTTBORN}\footnote{https://github.com/captain-exoplanet/misttborn} \citep{Mann2016, misttborn}. All the codes converge to a consistent solution for the stellar and BD parameters.

\begin{deluxetable}{ccc}
\tabletypesize{\footnotesize}
\tablewidth{0pt}

 \tablecaption{Comparison of parameters between analysis methods. The parameters here are the median values except for the {\tt EXOFASTv2} $R_p$, which shows both the median and the mode. \label{tab:compare}}

 \tablehead{
 \colhead{Parameter} & \colhead{{\tt SPC}/{\tt EXOFASTv2}} & \colhead{{\tt iSpec}/{\tt PARAM 1.3}}}

\startdata 
$M_\star$ ($\rm \mst$) & $1.80 \pm 0.06$ &  $1.78 \pm 0.02$\\
$R_\star$ ($\rm \rst$)&  $1.70 \pm 0.05$ &  $1.77 \pm 0.04$\\
$\log{g}$ &  $4.23 \pm 0.03$ &  $ 4.17 \pm 0.02$\\
$T_{\rm eff}$ (K) &  $7650 \pm 160$ &  $ 7639 \pm 105$\\
$[{\rm Fe/H}]$ & $0.30 \pm 0.09$ &  $ 0.61 \pm 0.07$\\
$[{\rm Ni/H}]$ & $-$ &  $ 0.58 \pm 0.09$\\
$[{\rm Ca/H}]$ & $-$ &  $ -0.40 \pm 0.11$\\
$[{\rm Sc/H}]$ & $-$ &  $ 0.10 \pm 0.14$\\
$[{\rm Mg/H}]$ & $-$ &  $ 0.25 \pm 0.15$\\
$v_{\rm rot} \sin{i_\star} $ (km/s) &  $28.6 \pm 0.4$ &  $25.0 \pm 0.3$\\
$P_{\rm rot}$ (days) & $3.01 \pm 0.09$ & $3.64 \pm 0.13$\\
Age (Gyr) & $0.18^{+0.17}_{-0.11}$ & $0.14 \pm 0.04$\\
\\
\hline\hline
Parameter & {\tt EXOFASTv2} & {\tt GeePea}/{\tt Systemic Console} \\
\hline
$M_b$ ($\rm M_J$) &  $53.7 \pm 1.2$ & $53.3 \pm 1.1$\\
$R_b$ ($\rm R_J$)&  $1.34 \pm 0.26$ & $1.28 \pm 0.29$\\
$R_{b, {\rm mode}}$ ($\rm R_J$)&  $1.27 \pm 0.15$ & $-$\\
Period (days) & $3.6772 \pm 0.0001$ & $3.6775 \pm 0.0002$\\
$a/R_*$ & $7.22 \pm 0.22$ &  $7.47 \pm 0.19$\\
$R_b/R_*$ & $0.0805 \pm 0.015$ &  $0.0724 \pm 0.015$\\
$b$ & $0.974 \pm 0.022$ & $0.956 \pm 0.023$\\
Inclination $i$ (degree) & $82.25 \pm 0.41$ & $82.65 \pm 0.38$\\
$e$ & $0$ (adopted) & $0$ (adopted)\\
\enddata
\vspace{-3.5cm}
\end{deluxetable}

\begin{figure*}[!ht]
\centering
\includegraphics[width=0.9\textwidth,height=1.07\textwidth]{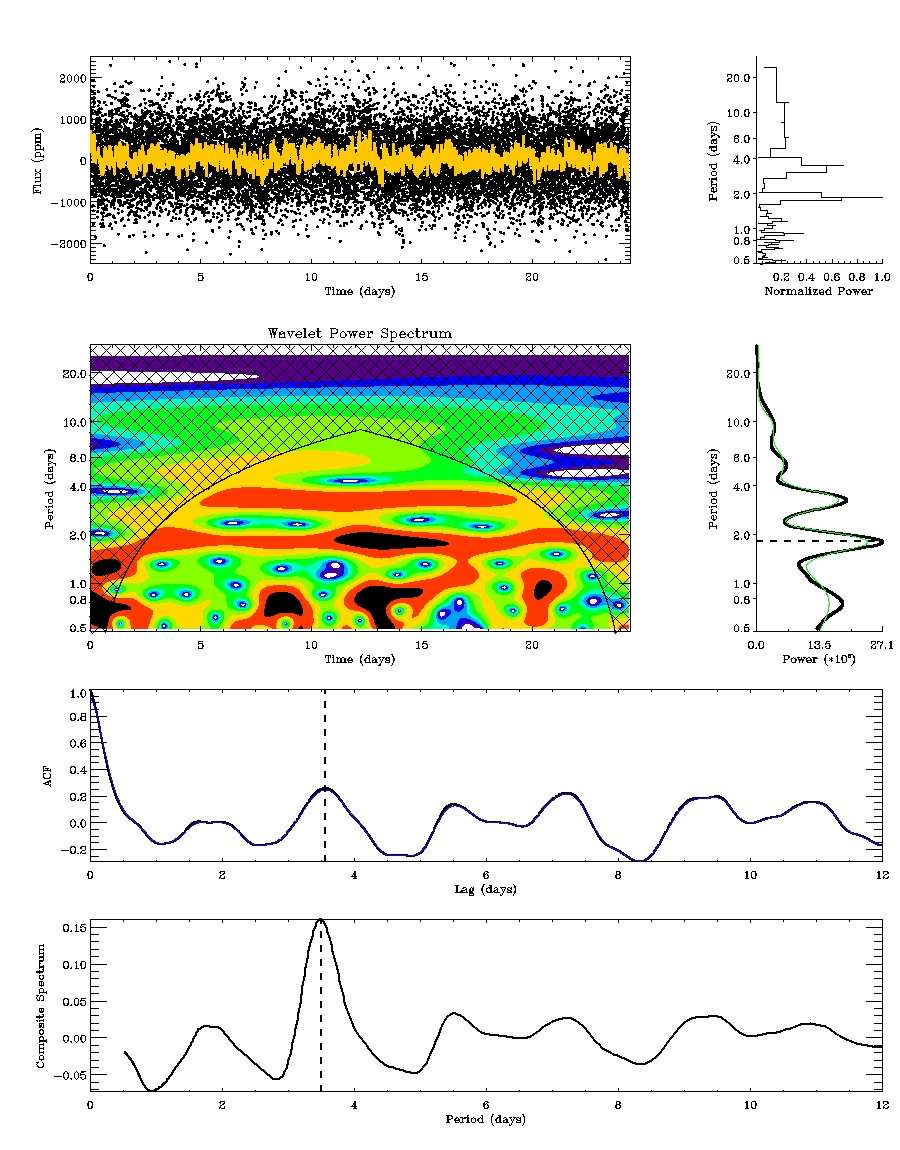}
\caption{The analysis of modulations in the light curve of TOI-503. Description of the images from top to bottom: First (left): The light curve cleared for primary and secondary transits. The orange points represent the light curve filtered with a boxcar function of 6 hours width. (right): LS periodogram with the power normalized to the power of the highest peak. Second (left): WPS as a function of the period and time. The different colors represent the strength of the power spectrum, where red and dark colors correspond to a higher power, and blue and light colors correspond to lower power. Hatched lines mask the zone of the diagram delimiting the cone of influence - the region where reliable rotation periods can be measured. (right): GWPS - the projection of the WPS on the Period axis (black line) with the corresponding Gaussian fit (green line). Third: ACF of the light curve. The dashed line points to the selected main periodicity. Fourth: CS of the light curve. The dashed line points to the selected periodicity.} \label{fig:rotation}
\end{figure*}

\begin{figure*}[!ht]
\centering
\includegraphics[width=1.0\textwidth,height=0.33\textwidth]{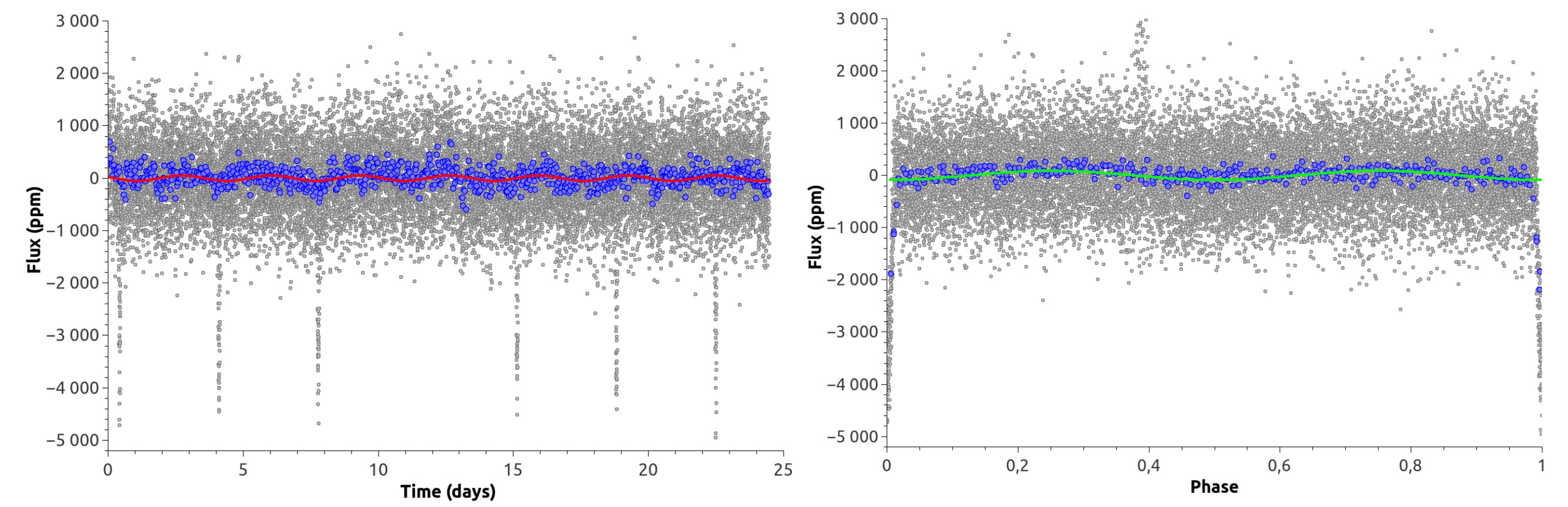}
\caption{The normalized light curve of TOI-503 observed by TESS is plotted in grey with blue points denoting the binned data. The left subplot shows the fit of the rotation signal (red curve), and the right subplot shows the phased data with the fit of the ellipsoidal deformation signal (green curve).} \label{fig:signals}
\end{figure*}

\subsubsection{Analysing the surface rotation}\label{sec:rotation}
Even though it is not completely clear that A-type stars have spots, there are a variety of studies on the discovery of spots on the well-known star, Vega \citep{Bohm2015, Petit2017, Balona2017} and more studies about the measurement of the rotation based on spot modulation for A-type stars \citep{Balona2011, Balona2013}. There is even previous evidence of the detectable presence of spots on Am-type stars \citep{Balona15}. So, it is reasonable to search for the signature of the rotation period through the modulation caused by the star spots in TOI-503. To do this, we use the SPOC two-minute cadence light curve of TOI-503. We removed the signal of the primary and secondary transits using the known ephemerides and then filled all gaps, including the transits and the data-transfer gaps, using in-painting techniques based on a multi-scale discrete cosine transform as described in \citet{Garcia2014} and \citet{Pires2015}.

We then search for modulation in the resulting light curve by performing the following steps. First, we perform a time-frequency analysis based on wavelets decomposition \citep{Torrence1998, Marthur2010, Garcia14} to compute the wavelet power spectrum (WPS), which we subsequently project on the period axis to form the global wavelets power spectrum (GWPS). In the second step, we perform auto-correlation function analysis \citep[ACF,][]{McQuillan2014} to extract the most significant signal, which corresponds to a particular period. Finally, by a combination of previous two steps (specifically multiplying them), we create a function called the Composite Spectrum (CS) \citep{Ceillier2016, Ceillier2017}. As these steps are sensitive to different types of artifacts in a light curve, by deriving the CS, we can mask such artifacts and highlight a periodic signal created by stellar activity such as star spots. The pipeline that combines these different techniques has been applied to simulated data \citep{Aigrain2015} and has been already performed to a large number of solar-like stars and red giants \citep[e.g.][]{Santos2019} with reliable success.

The original light curve analysis with the transits provides a period of $\rm P_{GWPS} = 3.66$ days, corresponding to the orbital period of the BD. Once the transits are removed, we find a period of $\rm P_{GWPS} = 3.24$ days with the wavelet analysis and $\rm P_{ACF} = 3.55$ days with the ACF analysis. The height of the peak in the ACF ($H_{ACF}$, measured from the maximum to the adjacent minima) is ~0.5, which fulfills our criteria for a reliable result ($H_{ACF>0.4}$). We note that we detect the overtone of half of the real rotation period $\rm P_{GWPS} = 1.8$ days using wavelets decomposition analysis, which can be seen in Figure \ref{fig:rotation}. This period is detected in power spectra quite often and happens when we observe active regions on the visible side of the star and diametrically opposite side of the star. This period is absent in the CS power spectrum and reveals the final period of $\rm P_{CS} = 3.50 \pm 0.12$ days. These results are slightly lower than the BD orbital period, but still quite close to it. The rotation period found (Table \ref{tab:compare}) also agrees with the values obtained spectroscopically from the {\tt iSpec} and {\tt SPC} analyses.

However, given the close values of the rotation period with the orbital period, we cannot completely rule out that the modulation we measure is affected by the orbital motion of the BD. Such an effect would project to LC in the form of additional signals, which are the ellipsoidal deformation signal on the primary star, and the optical reflection plus thermal phase curve signal. We try to distinguish these signals to find the real rotation period. To do so, we apply LS periodograms to the SPOC LC and fit the most significant periods by using the Levenberg-Marquardt minimization method. In Fig. \ref{fig:signals} we provide the ellipsoidal deformation signal with a period of half of the orbital period with maxima corresponding to phases 0.25 and 0.75 and rotational signal with a period of 3.3 days. We are not able to see any other signals in the LC. That is possibly linked with the nature of this BD. However, future investigation is needed to find an answer.

\subsubsection{Stellar parameters from Gaia DR2}
\begin{figure}[!ht]
\centering
\includegraphics[width=0.46\textwidth, trim = {0.0cm 0.0cm 0.0cm 0.0cm}]{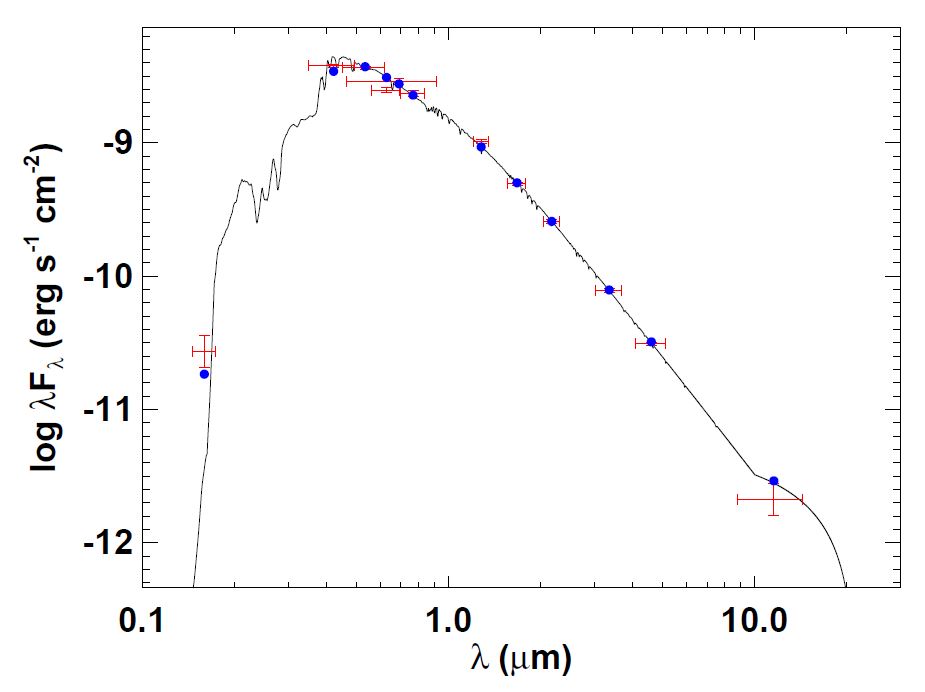}
\caption{SED fit using {\it Gaia\/} DR2 parallax with magnitudes from {\it Tycho-2} ($B_T V_T$), \citet{Paunzen:2015} (Str\"{o}mgren $ubvy$), APASS ($BVgri$), {\it 2MASS} ($JHK_S$), {\it WISE} (W1--W4), and the $G$ magnitude from {\it Gaia \/}. The SED measurements are in red with the model in blue. The point near $0.15\mu $m is from GALEX.} \label{fig:sed}
\end{figure}

As an additional independent check on the derived stellar parameters, we performed an analysis of the broadband spectral energy distribution (SED) together with the {\it Gaia\/} DR2 parallax in order to determine an empirical measurement of the stellar radius, following the procedures described in \citet{Stassun:2016,Stassun:2017,Stassun:2018}. We pulled the $B_T V_T$ magnitudes from {\it Tycho-2}, the Str\"{o}mgren $ubvy$ magnitudes from \citet{Paunzen:2015}, the $BVgri$ magnitudes from APASS, the $JHK_S$ magnitudes from {\it 2MASS}, the W1--W4 magnitudes from {\it WISE}, and the $G$ magnitude from {\it Gaia}. We also used the {\it GALEX} NUV and/or FUV fluxes, which are available in this case. Together, the available photometry spans the full stellar SED over the wavelength range 0.35--22~$\mu$m, and extends down to 0.15~$\mu$m when {\it GALEX} data are available (see Figure~\ref{fig:sed}). We performed a fit using Kurucz stellar atmosphere models, with the priors on effective temperature ($T_{\rm eff}$), surface gravity ($\log g$), and metallicity ([Fe/H]) from the spectroscopically determined values. The remaining free parameter is the extinction ($A_V$), which we restricted to the maximum line-of-sight value from the dust maps of \citet{Schlegel:1998}. 

The resulting fit is very good (Figure~\ref{fig:sed}) with a reduced $\chi^2$ of 4.8. The best fit extinction is $A_V = 0.00^{+0.06}_{-0.00}$, which is consistent with what we find with {\tt EXOFASTv2 ($A_V = 0.038^{+0.028}_{-0.026}$)}. This zero extinction is consistent with the maximum line-of-sight extinction from the 1-dimensional dust maps from \cite{Schlegel:1998} of 0.09 mag as well as the $A_V = 0.12 \pm 0.06$ value from \cite{AmoresLepine:2004}. Integrating the (unreddened) model SED gives the bolometric flux at Earth of $F_{\rm bol} = 4.18 \pm 0.15 \times 10^{-9}$ erg~s~cm$^{-2}$. Taking the $F_{\rm bol}$ and $T_{\rm eff}$ together with the {\it Gaia\/} DR2 parallax, adjusted by $+0.08$~mas to account for the systematic offset reported by \citet{StassunTorres:2018}, gives the stellar radius as $R = 1.66 \pm 0.05$~R$_\odot$. We note that when we do not account for this systematic offset (as with our values in Table \ref{tab:compare} and \ref{fig:mr_age}), we measure roughly $2.3\%$ larger radii for the star and BD. This difference does not affect our final conclusions about this system. Finally, estimating the stellar mass from the empirical relations of \citet{Torres:2010} gives $M = 1.90 \pm 0.11 M_\odot$, which with the radius gives the density $\rho = 0.58 \pm 0.06$ g~cm$^{-3}$. We find that this independent check on the stellar mass and radius agrees with the values shown in Table \ref{tab:compare}.

\subsection{Estimating the age of the TOI-503 system}\label{subsec:evolutionary_models}
We report an age of $180^{+170}_{-110}$ Myr for TOI-503 using the MIST models and {\tt EXOFASTv2}. We find this consistent with the Yonsei-Yale (YY) isochrone models \citep{yonseiyale}, from which we report an age of $200^{+200}_{-130}$ Myr. Both the MIST and YY isochrone grids are incorporated into the framework of {\tt EXOFASTv2}. A stellar mass track is interpolated from the grids for the MIST or YY isochrones, and from this, an age is estimated \citep{exofastv2}. We reiterate that the metallicity range of MIST isochrones is -5.0 $\leq$ [Fe/H] $\leq$ 0.5, which may influence the accuracy of the age estimate given that we measure a spectroscopic [Fe/H] of $0.6$ with {\tt iSpec}. The YY isochrones have a metallicity range of -3.29 $\leq$ [Fe/H] $\leq$ 0.78, which makes this set of isochrones better suited to this system. Still, we find the stellar and BD properties to be consistent between the two isochrone models.  

\begin{figure}[!ht]
\centering
\includegraphics[width=0.44\textwidth, trim= {1.0cm 0.0cm 0.0cm -0.5cm}]{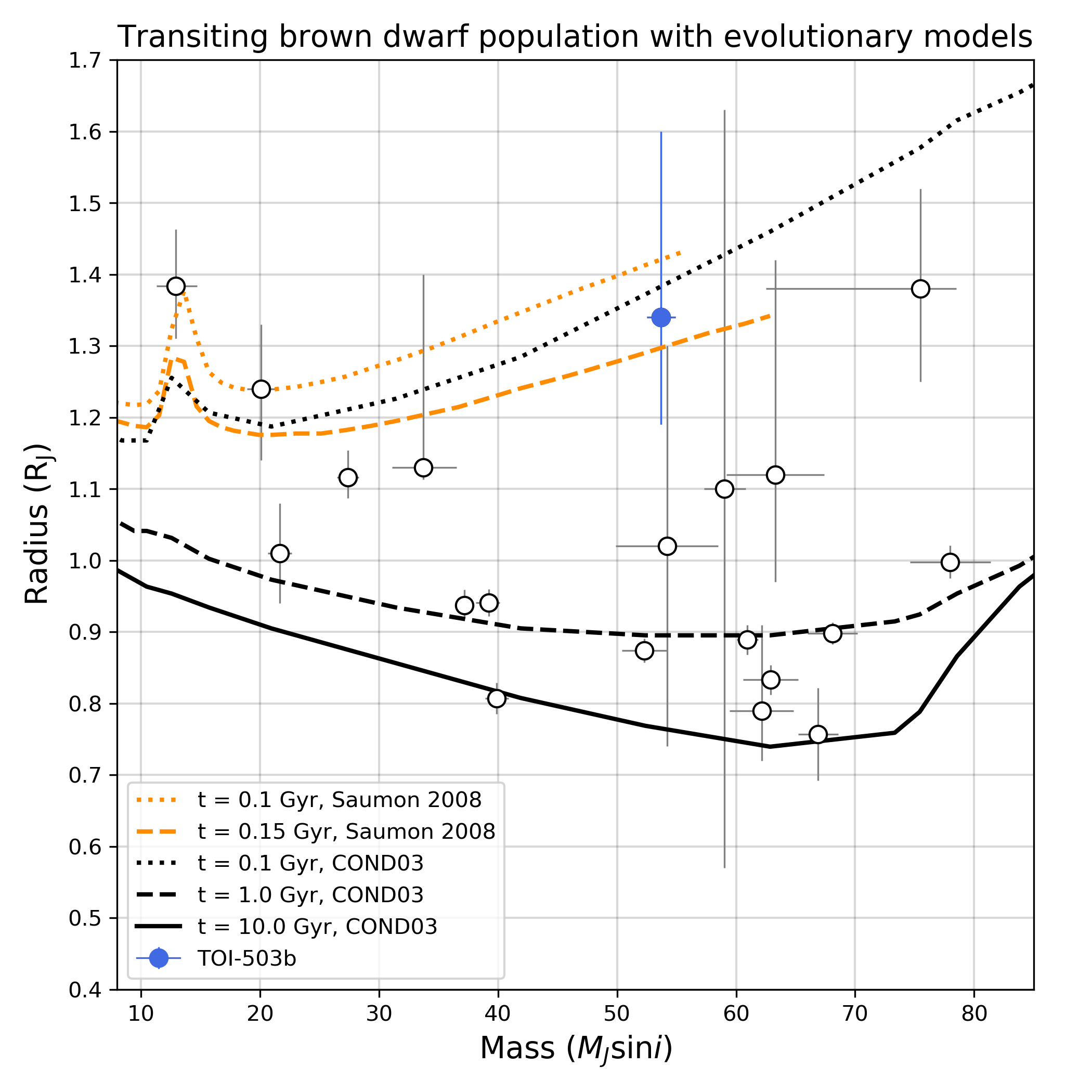}
\includegraphics[width=0.44\textwidth, trim= {1.0cm 0.0cm 0.0cm 0.0cm}]{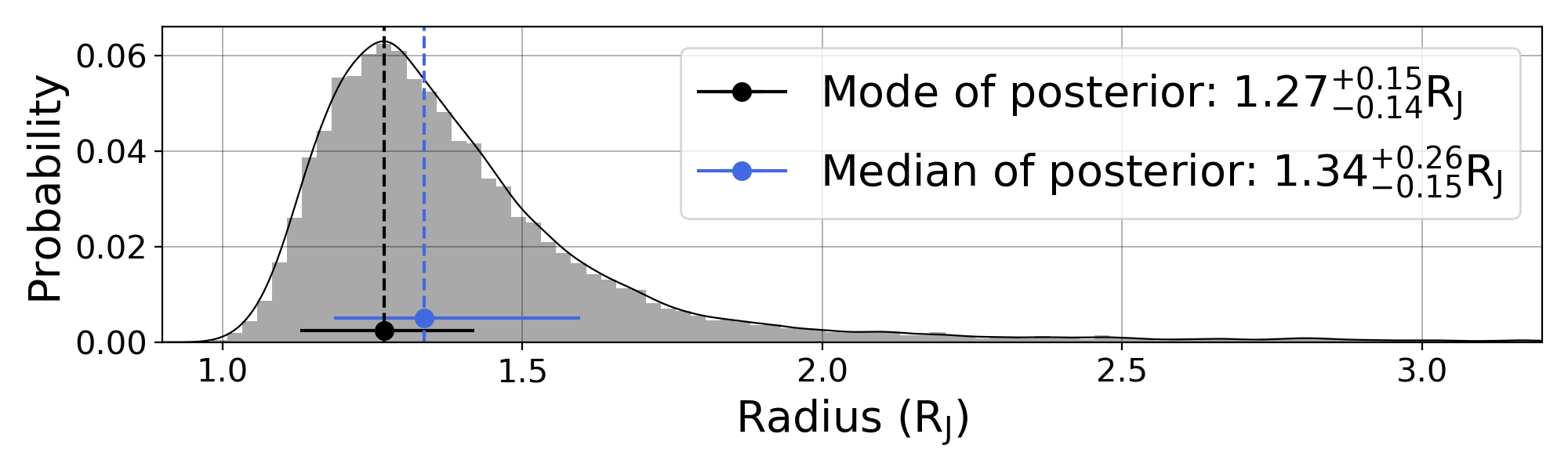}
\includegraphics[width=0.44\textwidth, trim= {1.0cm 0.0cm 0.5cm 0.0cm}]{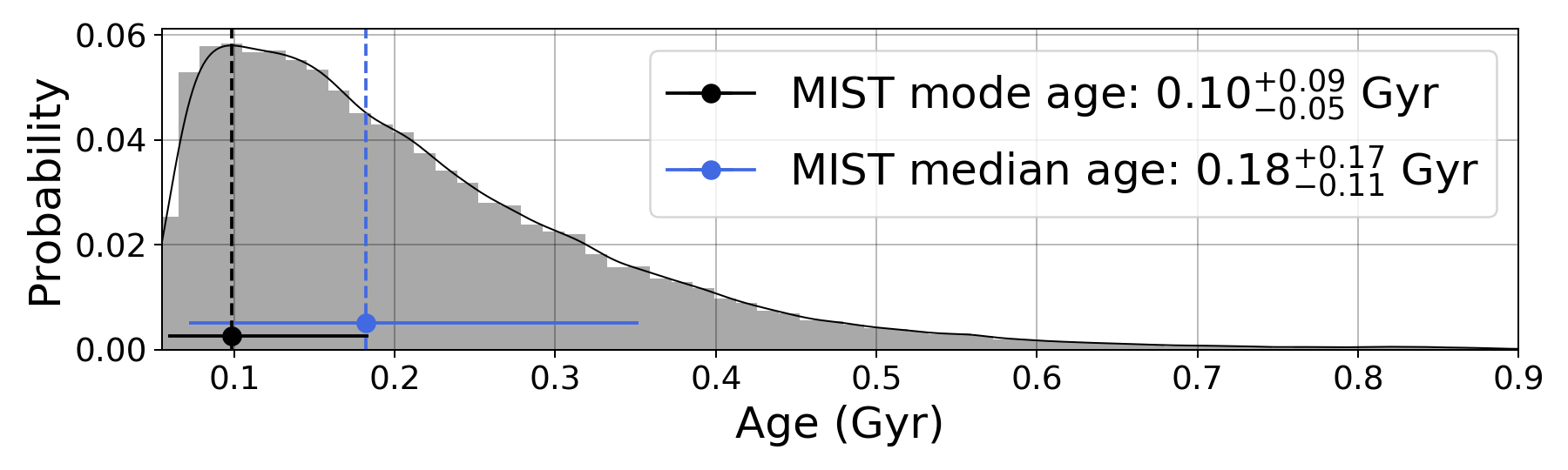}
\caption{Top: Evolutionary brown dwarf models of mass versus radius \citep{baraffe03, saumon08} with known transiting BDs over plotted. We use the \textit{median} results from {\tt EXOFASTv2} for the mass and radius for TOI-503b in this figure. Middle: Posterior distribution of the BD radius from the {\tt EXOFASTv2}/MIST results for TOI-503. The median value is reported in Table \ref{tab:exofast_table} as $1.34^{+0.26}_{-0.15}\rj$ and here, we report a value for the \textit{mode} of the posterior distribution to be $1.30^{+0.15}_{-0.14}\rj$. This is consistent with the posterior distribution for the BD radius using {\tt EXOFASTv2}/YY. Bottom: The posterior distribution for the age of TOI-503 showing the mode and median values for the age of the system.} \label{fig:mr_age}
\end{figure}

We now look at the \cite{baraffe03} (COND03) and \cite{saumon08} (SM08) substellar evolutionary models to examine how well they serve as predictors of the age of TOI-503b (Figure \ref{fig:mr_age}). The COND03 models present evolutionary tracks for irradiated giant planets and BDs, making them useful in the study of short-period BDs. The \cite{saumon08} models include details like metal-rich, metal-poor, and cloudy atmosphere models for low-mass stars and BDs, but do not include the effects of irradiation. However, both the COND03 and SM08 models are limited in their application to TOI-503b. The BD cooling models from \cite{baraffe03} indicate that an object with a mass on the order of $10\mj$ and an age greater than 500 Myr may maintain a radius of at least 1.0-1.2$\rj$ while in close proximity (semi-major axis of $a=0.046$AU) to a host star with $T_{\rm eff} = 6000$K. The difference between a non-irradiated and irradiated BD at ages up to 10 Gyr is roughly $0.1\rj$ \citep{baraffe03}. However, this is not the most appropriate comparison to TOI-503b primarily because of the large difference in the mass of the BD in this case ($53\mj$ versus $10\mj$ from the COND03 models), which means that TOI-503b has a higher internal luminosity that will affect its radius over time. We also expect a much hotter star at $T_{\rm eff}=7650$K (versus $T_{\rm eff}=6000$K) to have the effect of slowing the natural contraction of the BD's radius over time.

Lastly, given the grazing nature of the transit, which limits how well we may constrain the BD radius, we are limited in how thoroughly we may interpret what effects the BD mass and $T_{\rm eff}$ of the host star have on the radius of this substellar companion. This means that the COND03 models may only be used as a broad, qualitative check for the age of TOI-503b. The SM08 models should be treated in a similar, albeit less reliable, way as these do not consider the effects of irradiation. However, we are confident that this is one of the youngest intermediate-mass BDs ever found.

\section{Discussion}\label{sec:conclusion}

\subsection{The transiting BD population}
\begin{figure}[!ht]
\centering
\includegraphics[width=0.46\textwidth, trim = {0.0cm 0.0cm 0.0cm 0.0cm}]{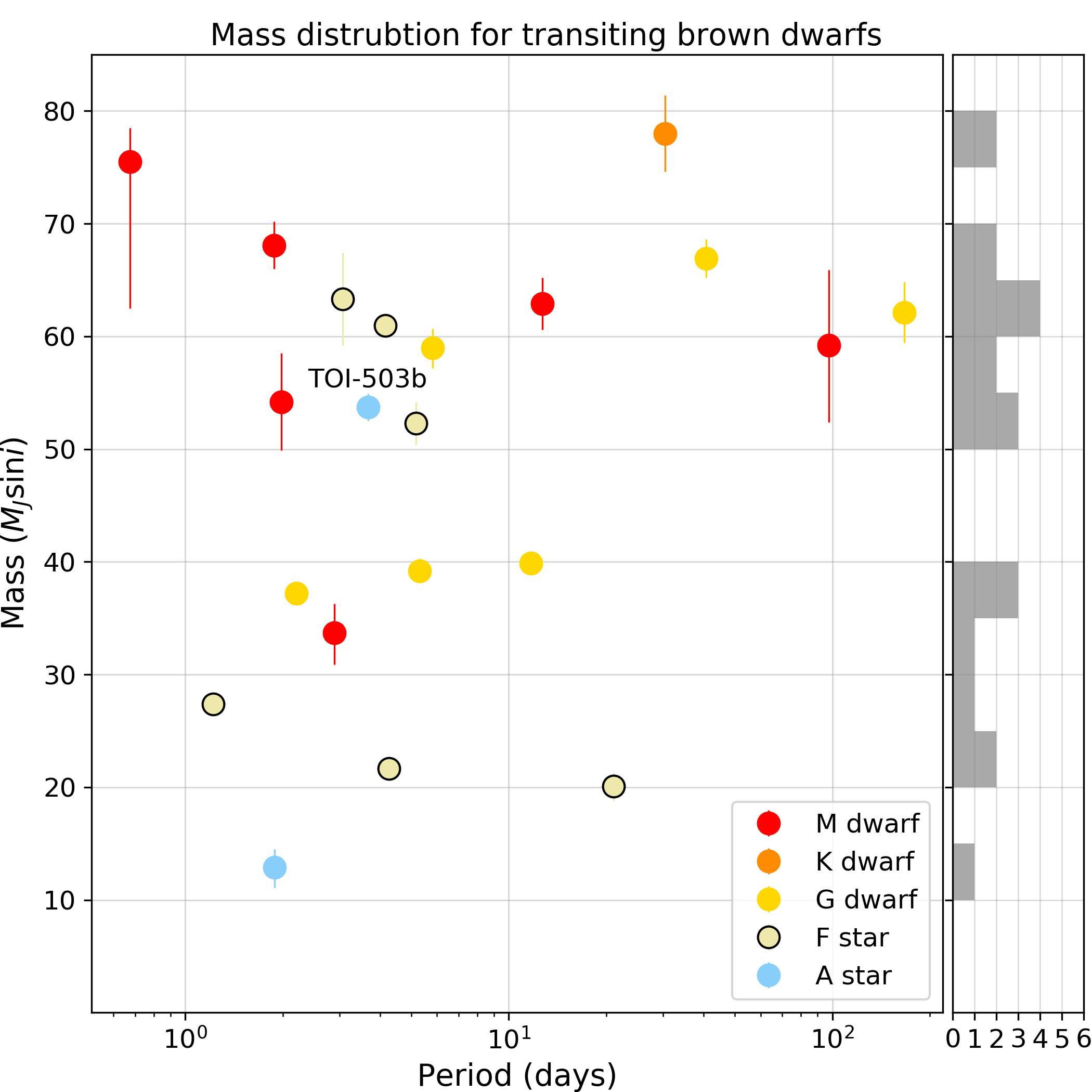}
\caption{The mass distribution over period for transiting BDs from Table \ref{tab:bdlist}. The color of each point indicates the spectral type of the star that hosts the BD. The histogram of the BD mass distribution is shown in the right panel with bin sizes of $5\mj$. The absence of BDs in the 40-50$\mj$ mass range can be seen here, but we caution that this may be a result of the small number of transiting BDs (21) that have been discovered to date.} \label{fig:mass_period}
\end{figure}

The mass-period diagram of transiting BDs (Figure \ref{fig:mass_period}) shows a sparse but diverse population. The total number of published transiting BDs, including TOI-503b in this work, is 21 (see Table \ref{tab:bdlist} for this list). The BD binary system, 2M0535-05 \citep{2M0535}, and the very young ($\sim$5-10 Myr) RIK 72b, which transits a pre-main sequence star \citep{david19_bd}, are not shown in Figure \ref{fig:mr_age} because their radii are above 3$\rj$ and do not correspond to the \cite{baraffe03} and \cite{saumon08} models, as discussed in Section \ref{subsec:evolutionary_models.} KOI-189b \citep{diaz14} has a mass of $78.0 \pm 3.4 \mj$ and is the most massive BD while HATS-70b, with a mass of $12.9 \pm 1.8 \mj$, is the least massive. This neatly places objects at the two extremes in mass of what is considered a BD, but \cite{diaz14} caution that KOI-189b may instead be a low-mass star. 

TOI-503b has an intermediate-mass of $53.7 \pm 1.2 \mj$, an inclination angle of $82.25^{+0.31}_{-0.41}$ degrees ($b=0.97\pm 0.02$), 
and adds to the diversity of objects found in the brown dwarf desert as it is one of the youngest BDs known to transit a main-sequence star. Past works have argued that there is a paucity of objects from 35-55 $M_J \sin{i}$, $P \leq 100$ days but this argument is difficult to support given the relatively small number of transiting BDs discovered and the fact that in recent years, 5 BDs (\cite{ad3116}, \cite{nowak17}, \cite{wasp128b}, \cite{carmichael19}, \cite{Persson19}, and this work) have been discovered in this intermediate mass-range, bringing new life, so to speak, to the desert. The recent growth in the discoveries of this type of BD could be a hint at an undisclosed population of intermediate-mass BDs in the BD desert. With the rise in the population of BDs in the intermediate mass-range, we can not rule out further reforesting of the driest region of the BD desert. However, we note that this is mostly a qualitative assessment of the distribution of intermediate-mass BDs, but an interesting feature to highlight nonetheless.

\subsection{Circularization timescales and orbital synchronization for TOI-503}
\label{sec:tides}

Based on our estimate of the age of TOI-503 (roughly 180 Myr) and the circular orbit of TOI-503b, we now consider the role tidal interactions have played in the orbital evolution of this system, namely, whether or not tides could have circularized the orbit of the BD. This comparison of circularization timescale to the system's age has implications for how the BD may have formed.

In order for a binary system affected by tides to be in a stable equilibrium, it must satisfy two conditions: the orbital angular momentum must be at least three times the sum of the rotational angular momenta of the two components, and the total angular momentum of the system must be greater than the critical value:
\begin{equation}
    L_{\mathrm{crit}}=4\bigg[\frac{G^2}{27}\frac{M_{\star}^3 M_{\mathrm{BD}}^3}{M_{\star}+M_{\mathrm{BD}}}(I_{\star}+I_{\mathrm{BD}})\bigg]^{\frac{1}{4}}
\end{equation}
where $I_{\star}$ and $I_{\mathrm{BD}}$ are the rotational moments of inertia of the star and BD, respectively \citep{hut1980}. We assume a value of $I_{\star}=\alpha_{\star} M_{\star}R_{\star}^2$ where $\alpha_{\star}=0.24$, interpolating the stellar moments of inertia from \cite{claret1989} to the mass of TOI-503. For the BD we assume the same internal structure as Jupiter, such that $\alpha_{\mathrm{BD}}=0.275$ \citep{Ni:2018}. We additionally assume that the orbit of the BD is well-aligned to the stellar rotation, i.e. $\sin i_{\star}\approx1$, in order to calculate the stellar rotation period from $v\sin i_{\star}$, and we assume the present-day stellar rotation rate for the quoted calculations. We find that $L_{\mathrm{tot}}=1.07 \pm 0.07 L_{\mathrm{crit}}$, and $L_{\mathrm{orb}}=5.0 \pm 0.5 L_{\mathrm{rot}}$. TOI-503 is thus Darwin stable; interestingly, the total angular momentum is consistent with being equal to the critical value, while the orbital angular momentum is close to twice the critical value relative to the rotational angular momentum, both much like KELT-1b \citep{kelt1b}.


From \cite{Jackson2008}, the timescale for orbital circularization timescale for a close-in companion is
\begin{equation}
\label{eqn:tidecirc}
    \frac{1}{\tau_e}=\bigg[\frac{63}{4}\sqrt{GM_{\star}^3}\frac{R_{\mathrm{BD}^5}}{Q_{\mathrm{BD}}M_{\mathrm{BD}}}+\frac{171}{16}\sqrt{G/M_{\star}}\frac{R_{\star}^5M_{\mathrm{BD}}}{Q_{\star}}\bigg]a^{-\frac{13}{2}}
\end{equation}
where $Q_{\star}$ and $Q_{\mathrm{BD}}$ are the tidal quality factors of the star and brown dwarf, respectively. \cite{Jackson2008} did not provide an expression for the tidal synchronization timescale, but \cite{Goldreich1966}, from whom \cite{Jackson2008} obtained their expressions, did. Rewriting this expression to use the terminology of this work, the synchronization timescale is
\begin{equation}
\label{eqn:tidesynch}
    \frac{1}{\tau_{\Omega}}=\frac{9}{4} \frac{G R_{\star}^3 M_{\mathrm{BD}}^2}{\alpha_{\star} M_{\star} Q_{\star} \Omega a^6}
\end{equation}
where $\Omega$ is the angular velocity of the star.

\begin{figure}[!ht]
\centering
\includegraphics[width=0.5\textwidth]{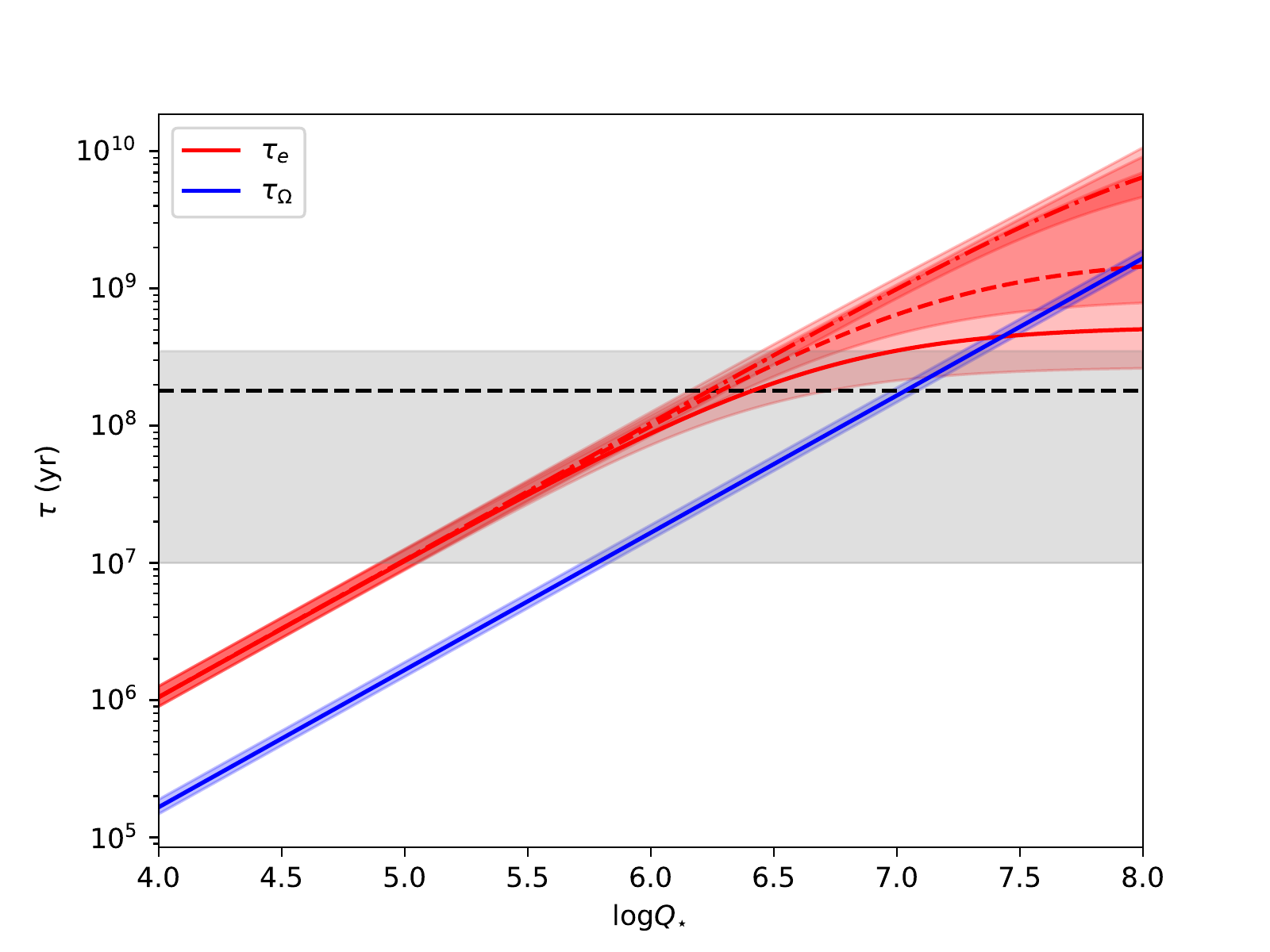}
\caption{Values of the tidal circularization timescale $\tau_e$ (red line; Eqn.~\ref{eqn:tidecirc}) and tidal synchronization timescale $\tau_{\Omega}$ (blue line; Eqn.~\ref{eqn:tidesynch}) as a function of stellar tidal quality factor $Q_{\star}$. The solid, dashed, and dot-dashed red lines correpond to values of the companion's quality factor $\log Q_{\mathrm{BD}}=4.5, 5, 6$, respectively; these lines flatten at large $Q_{\star}$ as the dissipation in the brown dwarf begins to dominate the system, while we only show these lines for $\tau_e$ as $\tau_{\Omega}$ does not depend on $Q_{\mathrm{BD}}$. The colored bands surrounding each line show the uncertainty on the timescale incorporating our measured uncertainties on the system parameters, \emph{but} assuming $Q_{\star}$, $Q_{\mathrm{BD}}$ fixed at the quoted values. The horizontal dashed line and gray region show the nominal system age and uncertainty thereon, respectively.} \label{fig:tidaltimescales}
\end{figure}

Tidal quality factors are difficult to measure. There are only three brown dwarfs with published constraints on $Q$, which indicate $\log Q_{\mathrm{BD}}>4.15-4.5$ \citep{Heller:2010,Beatty:2018}. Furthermore, there is disagreement in the literature about the values of $Q_{\star}$, with plausible values ranging from $10^4$ to $10^8$; furthermore, even for a single system the value of $Q_{\star}$ may change over time as the tidal forcing changes due to the orbital evolution of the system as well as stellar evolution \citep[e.g.,][]{Jackson2008,Penev2012}. Nonetheless, in order to assess the effect of the uncertain value of $Q_{\star}$ on the evolution of TOI-503, in Fig.~\ref{fig:tidaltimescales} we show the tidal damping timescales as a function of the tidal quality factors. As is apparent from the figure, the tidal timescales are shorter for the lower half of plausible values of $Q_{\star}$ and longer for the upper half. 

We provide preferring values $Q_{\star}=10^8$ and $Q_{\mathrm{BD}}=10^5$ following \cite{Persson19}. These values have been chosen using the observational quantification of the dissipation of the stellar equilibrium tide by \cite{CollierCameron2018} in hot-Jupiter systems and of the dissipation of tides in Jupiter by \cite{Lainey2009}, respectively. Although EPIC 212036875 \citep{Persson19} is less massive than TOI-503, \cite{CollierCameron2018} did not find any significant \teff\ dependence of $Q_{\star}$ for the stellar equilibrium tide, justifying this assumption. For the orbital period larger than half of the rotation period of the star tidal inertial waves (i.e., one of the components of the dynamical tide; \citep[e.g.,][]{Ogilvie2007}) can be excited in convective regions \citep{Bolmont2016}. However, for a massive star such as TOI-503, the convective envelope should be very thin leading to a negligible dissipation of tidal inertial waves \citep{Mathis2015, Gallet2017} while this dissipation can also be neglected in the convective core because this is a full sphere \citep[e.g.,][]{Ogilvie2004, Wu2005}. At the same time, the presence of this core may prevent an efficient dissipation of tidal gravity waves propagating in the radiative layers of TOI-503 \citep{Barker2010, Guillot2014}. Therefore, the estimate we provide here using the equilibrium tide values should be reasonable. We also note, that an alternative tidal model relying on dynamical tides within the radiative envelope of hot stars was presented by \cite{zahn1977}; using this model, which may be more appropriate for a hot star like TOI-503 predicts tidal damping timescales more than an order of magnitude larger than the \cite{Jackson2008} model for the largest plausible values of $Q_{\star}$. However, given this uncertainty in the appropriate tidal model and value of $Q_{\star}$, we cannot draw any firm conclusions on the tidal evolution of the system.

\subsection{TOI-503 in Context among Am Star Binaries}
Am stars are commonly found in binary systems \citep[e.g.,][]{Carquillat07}, and rotate more slowly than is typical for field A stars \citep[e.g.,][]{Abt95}. The Am nature of these stars is thought to be due to their slow rotation and it has been hypothesized that there may be a link between the binarity and slow rotation, but the exact mechanisms involved are still not well known \citep[e.g.,][]{BohmVitense2006}. It has also been noted that not all slowly-rotating A stars are Am stars \citep{Abt2009}. While many short-period Am binary systems could have experienced tidal synchronization, a significant number of Am binaries are too widely separated to experience significant tidal effects within their main sequence lifetimes \citep{Carquillat07}. Due to the systematic uncertainty on the tidal timescales for TOI-503 (\S\ref{sec:tides}), we cannot make any firm conclusions as to how any tidal braking experienced by the TOI-503 primary contributed to its nature as an Am star.

\begin{figure}[!ht]
\centering
\includegraphics[width=0.47\textwidth]{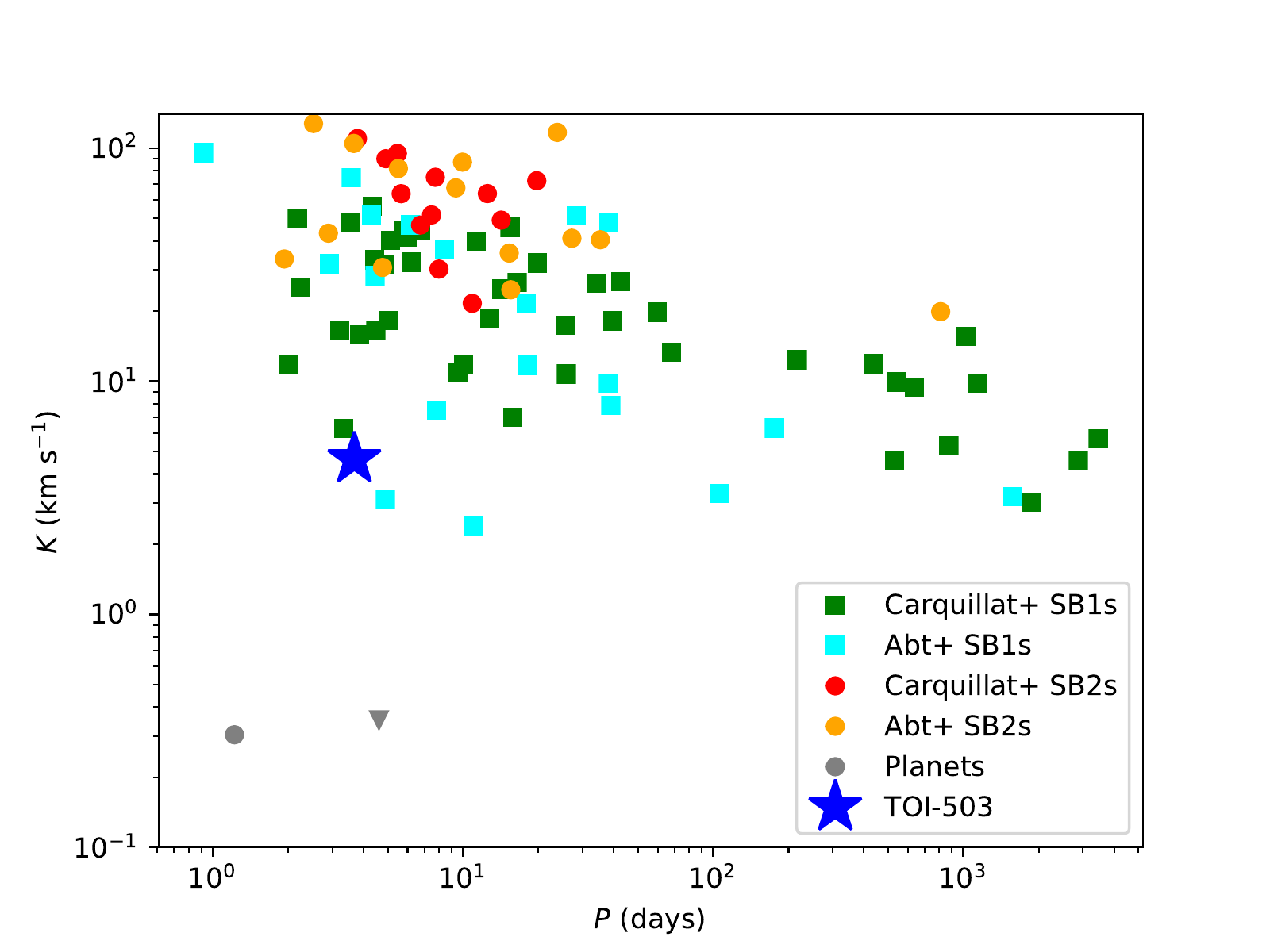}
\caption{TOI-503 in context among the population of Am star binaries from \cite{Abt1985}, \cite{Carquillat07}, and \cite{smalley14}, as well as the planets WASP-33b and KELT-19Ab \citep[the latter marked with a triangle as there is only an upper mass limit available;][]{Siverd2018}). The circles show SB2s where we have the RV semi-amplitude $\mathbf{K}$ for both components, while the squares show SB1s where we only have $\mathbf{K}$ for one component. TOI-503's RV semi-amplitude is among the smallest of any known Am star binary.}\label{fig:Ampopulation}
\end{figure}

Although it lies at the lower end of the envelope in mass ratio, the TOI-503 system does not otherwise stand out significantly from the known population of Am binaries. In Fig.~\ref{fig:Ampopulation} we show TOI-503 in context in the RV semi-amplitude-period plane for known Am star binaries. TOI-503 has among the lowest $K$ value of any known such system, but not the lowest. \cite{Boffin2010} showed that the mass ratio distribution of Am star binaries is uniform, and in this context the existence of TOI-503~b at a very small mass ratio is not surprising. Future surveys more sensitive to very small mass ratio Am binaries will be necessary to determine whether the mass ratio distribution eventually tails off.


Apart from TOI-503, other Am stars known to host a low mass companion ($M_b < 80\mj$), are WASP-33 \citep{wasp33b} and KELT-19A \citep{Siverd2018}, but the mass ratio $q$ for both the systems is even smaller than the TOI-503 system.

\subsection{How did TOI-503b form?}
The age of TOI-503 is approximately 180 Myr. In section 4.2, we have shown that given our consideration of the \cite{Jackson2008} and \cite{zahn1977} prescription of the tidal evolution of the system, we cannot say with certainty whether or not TOI-503 circularized the BD's orbit. Our interpretation of this is that TOI-503b may have formed at a larger orbital distance from its host star or also at a slightly larger eccentricity or more simply formed in-situ in a nearly circular orbit.

Now the question becomes which formation mechanism (core accretion or fragmentation) is more plausible. To address this point, we can look at some of the nearest neighbors, in terms of mass, to TOI-503b: AD 3116b ($M_b=54\mj$), EPIC 212036875b ($M_b=52\mj$), and CWW 89Ab ($M_b=39\mj$). As shown in two independent studies \citep{carmichael19, Persson19}, EPIC 212036875b is an eccentric ($e = 0.132$), short-period ($P=5.16$ days) BD that orbits an F-type star. The \cite{Persson19} study found EPIC 212036875b to likely have formed farther from its host star via gravitational disk instabilities and then quickly migrate to its current, close-in eccentric orbit. This is argued because core accretion is not effective to grow a $50\mj$ object \citep{kratter:2016} where fragmentation in the disk certainly could be and this may be the case for TOI-503b.

For AD 3116b, the circularization timescale is challenging to interpret, as \cite{ad3116} caution, given the mass ratio of this system ($q \approx 0.18$), the very short ($<2$ days) period, and the nature of the host M dwarf star. Though AD 3116b is young at a measured age of 600 Myr, given the nature of its orbit and host star, it is more difficult to infer a formation scenario. The narrative for CWW 89Ab is different still from these other examples, as \cite{cww89a} argue that it formed via core-accretion in part of a triple system that includes a wide secondary star, CWW 89B. From these few examples, we see that a difference in mass of 39$\mj$ (CWW 89Ab) versus 52$\mj$ (EPIC 212036875b) may have origins in different formation scenarios. This highlights how the mass of a short-period BD may not as strongly dictate a formation scenario as any single discovery might imply and that there are plausible non-in-situ formation pathways that come from both core accretion or disk fragmentation.

On the other hand, the in-situ formation scenario points to a possible way that Am stars can form. To our knowledge, this is the first time a BD has been detected orbiting an Am star in such a short-period. Detailed studies of Am stars report a binary fraction around 60-70\% \citep{Abt1985,Carquillat07}. In some such systems, the stellar companions are too distant for the tidal braking to be effective \citep[e.g.,][]{Siverd2018}. It possibly suggests that other processes may need to be invoked to explain their small rotational velocities. However, this may also be linked to the fact that it is relatively difficult to detect such a low mass companion as a BD around an Am star and our discovery of one such BD around an Am star in the BD desert may reflect this. However, further similar discoveries are needed to confirm if this is the correct explanation. The BD orbiting an Am star is a bridge connecting two areas that are not fully understood: the formation mechanisms and ultimate classification of BDs, and the creation, evolution, and behavior of Am stars. Such an overlap enables us to look at these areas from an entirely new perspective.

\section{Summary}
We have presented the analysis of the first BD known to transit an Am star, TOI-503b. This is the newest member of the brown dwarf desert and it orbits its host star in a circular, short-period ($P=3.67718 \pm 0.0001$ days) orbit. We measure the host star to have a mass of $M_\star = 1.80 \pm 0.06 \msun$, a radius of $R_\star = 1.70 \pm 0.05 \rsun$, an effective temperature of $T_{\rm eff} = 7650 \pm 160$K, and metallicity of $0.6\pm 0.1$ dex. The transit geometry of the system is grazing as revealed by the TESS light curve. The BD has a radius of $R_b = 1.34^{+0.26}_{-0.15}\rj$ and mass of $M_b = 53.7 \pm 1.2 \mj$, which places it in the driest part of the BD desert. The age of the system is estimated to be $\sim$180 Myr using MIST and YY isochrones. Given the difficulty in measuring the tidal interactions between brown dwarfs and their host stars, we cannot make any firm conclusions whether this brown dwarf formed in-situ or has had its orbit circularized by its host star. Instead, we offer an examination of plausible values for the tidal quality factor for the star and brown dwarf, and also provide the preferring value. \\

\section{Acknowledgements}

J.\v{S}. and P.K. would like to acknowledge the support from GACR international grant 17-01752J. J.\v{S}. would like to acknowledge the support from source 116-09/260441, Institute of Theoretical Physics, Charles University in Prague, Czech Republic. J.K., S.G., M.P., S.C., A.P.H., H.R., M.E. and K.W.F.L. acknowledge support by Deutsche Forschungsgemeinschaft (DFG) grants PA525/18-1, PA525/19-1, PA525/20-1, HA 3279/12-1, and RA 714/14-1 within the DFG Schwerpunkt SPP 1992, Exploring the Diversity of Extrasolar Planets. S.C. acknowledges the Hungarian Scientific Research Fund (OTKA) Grant KH-130372. This work is partly supported by JSPS KAKENHI Grant Numbers JP18H01265 and JP18H05439, and JST PRESTO Grant Number JPMJPR1775. S.M. acknowledges support by the Spanish Ministry through the Ramon y Cajal fellowship number RYC-2015-17697. R.A.G. acknowledges the support from the PLATO/CNES grant. M.S. acknowledges the Postdoc@MUNI project CZ.02.2.69/0.0/0.0/16-027/0008360. S. M. acknowledges support from the ERC SPIRE 647383 grant. M.F., C.M.P. and I.G. gratefully acknowledge the support of the  Swedish National Space Agency (DNR 163/16  and 174/18). K.G.S.\ acknowledges partial support from NASA grant 17-XRP17 2-0024. This paper includes data collected by the \textit{Kepler} mission. Funding for the \textit{Kepler} mission is provided by the NASA Science Mission directorate. Some of the data presented in this paper were obtained from the Mikulski Archive for Space Telescopes (MAST). STScI is operated by the Association of Universities for Research in Astronomy, Inc., under NASA contract NAS5-26555. Based on observations made with the Nordic Optical Telescope, operated by the Nordic Optical Telescope Scientific Association at the Observatorio del Roque de los Muchachos, La Palma, Spain, of the Instituto de Astrof\'isica de Canarias. Based in part on observations collected at the European Organisation for Astronomical Research in the Southern Hemisphere under ESO program P103.C-0449. TWC would like to acknowledge the effort of the observers who acquired the ground-based photometry at FLWO, LCO, CHAT, and FLI as part of the TESS Follow-up Program. Thanks to Alex J. Mustill for useful discussions. Thanks to Scott Gaudi for useful discussions. 
The PARAS spectrograph is fully funded and being supported by Physical Research Laboratory (PRL), which is part of Department of Space, Government of India. R.S. and A.C. would like to thank Director, PRL for his support and acknowledges the help from Vishal Shah and Mount Abu Observatory staff at the time of observations. A.C. is grateful to Suvrath Mahadevan from Pennsylvania University, USA and Arpita Roy from Caltech, USA for their tremendous efforts in the development of the PARAS data pipeline.

\bibliographystyle{aasjournal}
\bibliography{astro_citations}

\begin{deluxetable*}{cccccccccc}
\tabletypesize{\footnotesize}
\tablewidth{0pt}

 \tablecaption{List of published transiting brown dwarfs as of June 2019. \label{tab:bdlist}}

 \tablehead{
 \colhead{Name} & \colhead{$P$ (days)} & \colhead{$\rm M_{BD}/M_J$} & \colhead{$\rm R_{BD}/R_J$}& \colhead{e} & \colhead{$\rm M_\star/\mst$} &\colhead{$\rm R_\star/\rst$}& \colhead{$\rm T_{eff} (K)$}&\colhead{[Fe/H]} &\colhead{Reference}}
 \startdata 
 TOI-503b & 3.677 & $53.7 \pm 1.2$ & $1.34^{+0.26}_{-0.15}$ & 0 (adopted) & $1.80 \pm 0.06$ & $1.70 \pm 0.05$ & $7650 \pm 160$ & $+0.61 \pm 0.07$ & this work \\
 HATS-70b & 1.888 & $12.9^{+1.8}_{-1.6}$ & $1.38^{+0.08}_{-0.07}$ & $<0.18$ & $1.78 \pm 0.12$ & $1.88^{+0.06}_{-0.07}$ & $7930  ^{+630}_{-820}$ & $+0.04 \pm 0.11$ & 1\\
 KELT-1b & 1.218 & $27.4 \pm 0.9$ & $1.12 \pm 0.04$ & $0.01 \pm 0.01$ & $1.34 \pm 0.06$ & $1.47 \pm 0.05$ & $6516 \pm 49$ & $+0.05 \pm 0.08$ & 2\\
 NLTT 41135b & 2.889 & $33.7 \pm 2.8$ &  $1.13 \pm 0.27$ & $<0.02$ & $0.19 \pm 0.03$ & $0.21 \pm 0.02$ & $3230 \pm 130$ & $-0.25 \pm 0.25$ & 3\\
 LHS 6343c & 12.713 & $62.9 \pm 2.3$ & $0.83 \pm 0.02$ & $0.056 \pm 0.032$ & $0.37\pm 0.01$ & $0.38\pm 0.01$ & - & $+0.02 \pm 0.19$ & 4\\
 LP 261-75b & 1.882 & $68.1 \pm 2.1$ & $0.90 \pm 0.02$ & $<0.007$ & $0.30 \pm 0.02$ & $0.31 \pm 0.01$ & $3100 \pm 50$ & - & 5\\
 WASP-30b & 4.157 & $61.0 \pm 0.9$ & $0.89 \pm 0.02$ & 0 (adopted) & $1.17 \pm 0.03$ & $1.30 \pm 0.02$ & $6201 \pm 97$ & $-0.08 \pm 0.10$ & 6\\
 WASP-128b & 2.209 & $37.2 \pm 0.9$ & $0.94 \pm 0.02$ & $<0.007$ & $1.16 \pm 0.04$ & $1.15 \pm 0.02$ & $5950 \pm 50$ & $+0.01 \pm 0.12$ & 7\\
 CoRoT-3b & 4.257 & $21.7 \pm 1.0$ & $1.01 \pm 0.07$ & 0 (adopted) & $1.37 \pm 0.09$ & $1.56 \pm 0.09$ & $6740 \pm 140$ & $-0.02 \pm 0.06$ & 8\\
 CoRoT-15b & 3.060 & $63.3 \pm 4.1$ & $1.12 \pm 0.30$ & 0 (adopted) & $1.32 \pm 0.12$ & $1.46 \pm 0.31$ & $6350 \pm 200$ & $+0.10 \pm 0.20$ & 9\\
 CoRoT-33b & 5.819 & $59.0 \pm 1.8$ & $1.10 \pm 0.53$ & $0.070 \pm 0.002$ & $0.86 \pm 0.04$ & $0.94 \pm 0.14$ & $5225 \pm 80$ & $+0.44 \pm 0.10$ & 10\\
 Kepler-39b & 21.087 & $20.1 \pm 1.3$ & $1.24 \pm 0.10$ & $0.112 \pm 0.057$ & $1.29 \pm 0.07$ & $1.40 \pm 0.10$ & $6350 \pm 100$ & $+0.10 \pm 0.14$ & 11\\
 KOI-189b & 30.360 & $78.0 \pm 3.4$ & $1.00 \pm 0.02$ & $0.275 \pm 0.004$ & $0.76 \pm 0.05$ & $0.73 \pm 0.02$ & $4952 \pm 40$ & $-0.07 \pm 0.12$ & 12\\
 KOI-205b & 11.720 & $39.9 \pm 1.0$ & $0.81 \pm 0.02$ & $<0.031$ & $0.92 \pm 0.03$ & $0.84 \pm 0.02$ & $5237 \pm 60$ & $+0.14 \pm 0.12$ & 13\\
 KOI-415b & 166.788 & $62.1 \pm 2.7$ & $0.79 \pm 0.12$ & $0.689 \pm 0.001$ & $0.94 \pm 0.06$ & $1.15 \pm 0.15$ & $5810 \pm 80$ & $-0.24 \pm 0.11$ & 14\\
 EPIC 201702477b & 40.737 & $66.9 \pm 1.7$ & $0.76 \pm 0.07$ & $0.228 \pm 0.003$ & $0.87 \pm 0.03$ & $0.90 \pm 0.06$ & $5517 \pm 70$ & $-0.16 \pm 0.05$ & 15\\
 EPIC 212036875b & 5.170 & $52.3 \pm 1.9$ & $0.87 \pm 0.02$ & $0.132 \pm 0.004$ & $1.29 \pm 0.07$ & $1.50 \pm 0.03$ & $6238 \pm 60$ & $+0.01 \pm 0.10$ &  18, 21 \\
 AD 3116b & 1.983 & $54.2 \pm 4.3$ & $1.02 \pm 0.28$ & $0.146 \pm 0.024$ & $0.28 \pm 0.02$ & $0.29 \pm 0.08$ & $3200 \pm 200$ & $+0.16 \pm 0.10$ & 17 \\
 CWW 89Ab & 5.293 & $39.2 \pm 1.1$ & $0.94 \pm 0.02$ & $0.189 \pm 0.002$ & $1.10 \pm 0.05$ & $1.03 \pm 0.02$ & $5755 \pm 49$ & $+0.20 \pm 0.09$ & 16, 18 \\
 RIK 72b & 97.760 & $59.2 \pm 6.8$ & $3.10 \pm 0.31$ & $0.146 \pm 0.012$ & $0.44 \pm 0.04$ & $0.96 \pm 0.10$ & $3349 \pm 142$ & - & 19\\
 NGTS-7Ab & 0.676 & $75.5^{+3.0}_{-13.7}$ & $1.38^{+0.13}_{-0.14}$ & 0 (adopted) & $0.48\pm 0.13$ & $0.61\pm 0.06$ & $3359\pm 106$ & - & 20\\
 2M0535-05a & 9.779 & $56.7 \pm 4.8$ & $6.50 \pm 0.33$ & $0.323 \pm 0.006$ & - & - & - & - & 22\\
 2M0535-05b & 9.779 & $35.6 \pm 2.8$ & $5.00 \pm 0.25$ & $0.323 \pm 0.006$ & - & - & - & - & 22\\
 \enddata 
 \tablecomments{References: 1 - \cite{zhou19}, 2 - \cite{kelt1b}, 3 - \cite{irwin10}, 4 -  \cite{johnson11_bd}, 5 - \cite{irwin18}, 6 - \cite{wasp30b}, 7 - \cite{wasp128b}, 8 - \cite{corot3b}, 9 - \cite{corot15b}, 10 - \cite{corot33b}, 11 - \cite{kepler39}, 12 - \cite{diaz14}, 13 - \cite{diaz13}, 14 - \cite{moutou13}, 15 - \cite{bayliss16}, 16 - \cite{nowak17}, 17 - \cite{ad3116}, 18 - \cite{carmichael19}, 19 - \cite{david19_bd}, 20 - \cite{jackman2019}, 21 - \cite{Persson19}}, 22 - \cite{2M0535}.
\end{deluxetable*}

\begin{deluxetable*}{ccccc}

\tabletypesize{\footnotesize}
\tablewidth{0pt}

 \tablecaption{Additional information on published transiting brown dwarfs. \label{tab:bdlist_magnitudes}}

 \tablehead{
 \colhead{Name} & \colhead{$\alpha_{\rm J2000}$} & \colhead{$\delta_{\rm J2000}$} & \colhead{V (magnitude)}& \colhead{Reference}}

 \startdata 
 TOI-503 & 08 17 16.89 & 12 36 04.76 & 9.40 & this work\\
 LP 261-75  & 09 51 04.58 & +35 58 09.47 & 15.43 & \cite{irwin18}\\
 NLTT 41135 & 15 46 04.30 & +04 41 30.06 & 18.00 & \cite{irwin10}\\
 LHS 6343   & 19 10 14.28 & +46 57 24.11 & 13.39 & \cite{johnson11_bd}\\
 KELT-1     & 00 01 26.92 & +39 23 01.70 & 10.70 & \cite{kelt1b}\\
 HATS-70    & 07 16 25.08 & $-$31 14 39.86 & 12.57 & \cite{hats70b}\\
 WASP-30    & 23 53 38.03 & $-$10 07 05.10 & 12.00 & \cite{wasp30b}\\
 WASP-128   & 11 31 26.10 & $-$41 41 22.30 & 12.50 & \cite{wasp128b}\\
 CoRoT-3    & 19 28 13.26 & +00 07 18.70 & 13.29 & \cite{corot3b}\\
 CoRoT-15   & 06 28 27.82 & +06 11 10.47 & 16.00 & \cite{corot15b}\\
 CoRoT-33   & 18 38 33.91 & +05 37 28.97 & 14.70 & \cite{corot33b}\\
 Kepler-39  & 19 47 50.46 & +46 02 03.49 & 14.47 & \cite{bouchy11}\\
 KOI-189    & 18 59 31.19 & +49 16 01.17 & 14.74 & \cite{diaz14}\\
 KOI-205    & 19 41 59.20 & +42 32 16.41 & 14.85 & \cite{diaz13}\\
 KOI-415    & 19 33 13.45 & +41 36 22.93 & 14.34 & \cite{moutou13}\\
 EPIC 201702477 & 11 40 57.79 & +03 40 53.70 & 14.57 & \cite{bayliss16}\\
 EPIC 212036875 & 08 58 45.67 & +20 52 08.73 & 10.95 & \cite{Persson19}\\
 CWW 89A    & 19 17 34.04 & $-$16 52 17.80 & 12.54 & \cite{nowak17}\\
 AD 3116    & 08 42 39.43 & +19 24 51.90 & 18.73 & \cite{ad3116}\\
 NGTS-7A & 23 30 05.20 & -38 58 11.71 & 15.50 & \cite{jackman2019}\\
 RIK 72  & 16 03 39.22 & $-$18 51 29.72 & 16.01 & \cite{david19_bd} \\
 2M0535-05$\rm ^a$ & 05 35 21.85 & $-$05 46 08.56 & 18.94G$\rm ^b$ & \cite{2M0535}\\
 \enddata
 \tablecomments{a -- The 2M0535-05 system is a brown dwarf binary, b -- G-band magnitude from the {\it Gaia \/} mission}
\end{deluxetable*}

\begin{deluxetable*}{lcccccc}
\tablecaption{Median values and 68\% confidence interval for TOI-503, created using {\tt EXOFASTv2} commit number 65aa674.  \label{tab:exofast_table}}
\tablehead{\colhead{~~~Parameter} & \colhead{Units} & \multicolumn{5}{c}{Values}}
\startdata
\smallskip\\\multicolumn{2}{l}{Stellar Parameters:}&\smallskip\\
~~~~$M_*$\dotfill &Mass (\msun)\dotfill &$1.80^{+0.06}_{-0.06}$\\
~~~~$R_*$\dotfill &Radius (\rsun)\dotfill &$1.70^{+0.05}_{-0.04}$\\
~~~~$L_*$\dotfill &Luminosity (\lsun)\dotfill &$8.96^{+0.54}_{-0.54}$\\
~~~~$\rho_*$\dotfill &Density (cgs)\dotfill &$0.51^{+0.04}_{-0.05}$\\
~~~~$\log{g}$\dotfill &Surface gravity (cgs)\dotfill &$4.23^{+0.03}_{-0.03}$\\
~~~~$T_{\rm eff}$\dotfill &Effective Temperature (K)\dotfill &$7650^{+140}_{-160}$\\
~~~~$[{\rm Fe/H}]$\dotfill &Metallicity (dex)\dotfill &$0.30^{+0.08}_{-0.09}$\\
~~~~$Age$\dotfill &Age (Gyr)\dotfill &$0.18^{+0.17}_{-0.11}$\\
~~~~$EEP$\dotfill &Equal Evolutionary Point \dotfill &$292^{+22}_{-31}$\\
~~~~$A_V$\dotfill &V-band extinction (mag)\dotfill &$0.038^{+0.028}_{-0.026}$\\
~~~~$\sigma_{SED}$\dotfill &SED photometry error scaling \dotfill &$3.9^{+1.7}_{-0.99}$\\
~~~~$\varpi$\dotfill &Parallax (mas)\dotfill &$3.878^{+0.059}_{-0.058}$\\
~~~~$d$\dotfill &Distance (pc)\dotfill &$257.9^{+3.9}_{-3.8}$\\
\smallskip\\\multicolumn{2}{l}{Brown Dwarf Parameters:}&b\smallskip\\
~~~~$P$\dotfill &Period (days)\dotfill &$3.67718\pm0.00010$\\
~~~~$R_b$\dotfill &Radius (\rj)\dotfill &$1.34^{+0.26}_{-0.15}$\\
~~~~$T_C$\dotfill &Time of conjunction (\bjdtdb)\dotfill &$2458492.05383\pm0.00053$\\
~~~~$T_0$\dotfill &Optimal conjunction Time (\bjdtdb)\dotfill &$2458506.76256\pm0.00039$\\
~~~~$a$\dotfill &Semi-major axis (AU)\dotfill &$0.05727\pm0.00063$\\
~~~~$i$\dotfill &Inclination (Degrees)\dotfill &$82.25^{+0.31}_{-0.41}$\\
~~~~$T_{eq}$\dotfill &Equilibrium temperature (K)\dotfill &$2011^{+27}_{-28}$\\
~~~~$M_b$\dotfill &Mass (\mj)\dotfill &$53.7\pm1.2$\\
~~~~$K$\dotfill &RV semi-amplitude (m/s)\dotfill &$4640^{+30}_{-27}$\\
~~~~$logK$\dotfill &Log of RV semi-amplitude \dotfill &$3.6673^{+0.0028}_{-0.0026}$\\
~~~~$R_b/R_*$\dotfill &Radius of planet in stellar radii \dotfill &$0.0805^{+0.015}_{-0.0090}$\\
~~~~$a/R_*$\dotfill &Semi-major axis in stellar radii \dotfill &$7.22^{+0.20}_{-0.22}$\\
~~~~$\delta$\dotfill &Transit depth (fraction)\dotfill &$0.0065^{+0.0028}_{-0.0014}$\\
~~~~$Depth$\dotfill &Flux decrement at mid transit \dotfill &$0.00452^{+0.00026}_{-0.00023}$\\
~~~~$\tau$\dotfill &Ingress/egress transit duration (days)\dotfill &$0.03836^{+0.00060}_{-0.00057}$\\
~~~~$T_{14}$\dotfill &Total transit duration (days)\dotfill &$0.0767^{+0.0012}_{-0.0011}$\\
~~~~$b$\dotfill &Transit Impact parameter \dotfill &$0.974^{+0.022}_{-0.015}$\\
~~~~$\delta_{S,3.6\mu m}$\dotfill &Blackbody eclipse depth at 3.6$\mu$m (ppm)\dotfill &$700^{+320}_{-160}$\\
~~~~$\delta_{S,4.5\mu m}$\dotfill &Blackbody eclipse depth at 4.5$\mu$m (ppm)\dotfill &$860^{+390}_{-200}$\\
~~~~$\rho_b$\dotfill &Density (cgs)\dotfill &$27^{+13}_{-11}$\\
~~~~$logg_P$\dotfill &Surface gravity \dotfill &$4.87^{+0.12}_{-0.17}$\\
~~~~$M_P\sin i$\dotfill &Minimum mass (\mj)\dotfill &$53.2\pm1.2$\\
~~~~$M_P/M_*$\dotfill &Mass ratio \dotfill &$0.02844^{+0.00039}_{-0.00037}$\\
~~~~$c_{1}$\dotfill &linear limb-darkening coeff \dotfill &$0.146^{+0.049}_{-0.050}$\\
~~~~$c_{2}$\dotfill &quadratic limb-darkening coeff \dotfill &$0.333^{+0.049}_{-0.048}$\\
\smallskip\\\multicolumn{2}{l}{Telescope Parameters:}&FIES&Ond\v{r}ejov&PARAS&TRES&Tautenburg\smallskip\\
~~~~$\gamma_{\rm rel}$\dotfill &Relative RV Offset (m/s)\dotfill &$29468^{+22}_{-20}$&$3549^{+77}_{-78}$&$-31^{+85}_{-84}$&$4571^{+21}_{-23}$&$-2766^{+43}_{-35}$\\
~~~~$\sigma_J$\dotfill &RV Jitter (m/s)\dotfill &$46^{+41}_{-30}$&$120\pm120$&$230^{+100}_{-66}$&$40^{+43}_{-40}$&$0.00^{+93}_{-0.00}$\\
~~~~$\sigma_J^2$\dotfill &RV Jitter Variance \dotfill &$2200^{+5500}_{-1900}$&$13000^{+40000}_{-22000}$&$53000^{+57000}_{-26000}$&$1600^{+5300}_{-2200}$&$-1800^{+11000}_{-3500}$\\
\smallskip\\\multicolumn{2}{l}{Transit Parameters:}&TESS UT oi50-3.-TE (TESS)\smallskip\\
~~~~$\sigma^{2}$\dotfill &Added Variance \dotfill &$-0.0000000323^{+0.0000000064}_{-0.0000000063}$\\
~~~~$F_0$\dotfill &Baseline flux \dotfill &$0.9999834\pm0.0000059$\\
\enddata
\end{deluxetable*}

\end{document}